\documentclass[conference]{IEEEtran}

\usepackage{booktabs} 

\usepackage{algpseudocode}
\usepackage{algorithm}
\usepackage{listings}
\usepackage{subfig}
\usepackage{footnote}
\usepackage{graphicx}
\usepackage{multirow}
\usepackage{lipsum}
\usepackage{amsmath}
\usepackage{xcolor}
\usepackage{soul}

\usepackage{amssymb}
\usepackage{pifont}
\newcommand{\cmark}{\ding{51}}%
\newcommand{\xmark}{\ding{55}}%

\newcommand{\codename}{{\tt DeClassifier}}
\newcommand{\objectAnalysis} {Object Layout Analysis}

\newcommand{\identificationOfCtorDtor} {Correct identification of Constructors and Destructors}

\newcommand{\cht} {CHT }
\renewcommand{\paragraph}[1]{\medskip\noindent{\bf{#1.}}}
\newcommand{\specialcell}[2][c]{%
	\begin{tabular}[#1]{@{}c@{}}#2\end{tabular}}

\ifCLASSINFOpdf

\else

\fi

\begin{document}

	\title{{\tt DeClassifier: } Class-Inheritance Inference Engine for Optimized C++ Binaries}
	
	\author{\IEEEauthorblockN{Rukayat Ayomide Erinfolami}
		\IEEEauthorblockA{Binghamton University\\
			rerinfo1@binghamton.edu}
		
		\and
		\IEEEauthorblockN{Aravind Prakash}
		\IEEEauthorblockA{Binghamton University\\
			aprakash@binghamton.edu}}

	\maketitle

	\begin{abstract}
		Recovering class inheritance from C++ binaries has several security benefits including problems such as decompilation and program hardening.
		Thanks to the optimization guidelines prescribed by the C++ standard, commercial C++ binaries tend to be optimized. 
		While state-of-the-art class inheritance inference solutions are effective in dealing with unoptimized code, their efficacy is impeded by optimization.
		Particularly, constructor inlining--or worse exclusion-- due to optimization render class inheritance recovery challenging. Further, while modern solutions such as MARX can successfully group classes within an inheritance sub-tree, they fail to establish directionality of inheritance, which is crucial for security-related applications (e.g. decompilation).
  We implemented a prototype of \codename\ using Binary Analysis Plattform (BAP) and evaluated \codename\ against 16 binaries compiled using gcc under multiple optimization settings. 
		We show that (1) \codename\ can recover 94.5\% and 71.4\% true positive directed edges in the class hierarchy tree under O0 and O2 optimizations respectively, (2) a combination of ctor+dtor analysis provides much better inference than ctor only analysis.

	\end{abstract}
	
	\section{Introduction}\label{sec:intro}
	Recovery of class inheritance of a C++ program is useful in many ways, and often necessary. 
	While extracting class hierarchy from source code is straightforward (e.g., class-hierarchy analysis ~\cite{vtvgcc:2012:tice,jang:2014:CFI:Src,haller2015shrinkwrap,bounov:2016:ivt,lee:2015:caver,zhang2016vtrust,Burow2017CFIXXOT}), recovering class hierarchy from a binary is hard~\cite{Sabanal:2007:CppRev,igor:ida:decompiler}, but useful. 
	For example, any attempt at C++ program decompilation must infer at least a partial class hierarchy from a binary~\cite{fokin2010reconstruction,fokin:2011:C++Rev}. 
	Similarly, defenses that enforce strict control-flow integrity (CFI) policies on C++ binaries rely on class hierarchy analysis (e.g., Marx~\cite{marx:andre}, VCI~\cite{vci:2017}). 
	
	Although RunTime Type Information (RTTI), a per-class type-revealing data structure may be present in certain C++ programs, it is often absent in COTS binaries.  
	On the one hand, RTTI structure contains information about the parents of a given polymorphic class, which can be used to reliably reconstruct the class hierarchy of a program.
	But on the other hand, use of RTTI is discouraged in commercial code due to the high runtime overhead imposed by operators (i.e., {\tt dynamic\_cast} and {\tt typeinfo}) that use RTTI.
	
	In fact, most commercial-off-the-shelf (COTS) are closed source and do not contain RTTI in the binary.
	Without RTTI, inferring class hierarchy (inferring high level semantics in general) from COTS C++ software poses multiple challenges. 
	First, most solutions (e.g., VCI~\cite{vci:2017}, SmartDec~\cite{fokin:2011:C++Rev,fokin2010reconstruction}, HexRays Decompiler~\cite{igor:ida:decompiler}) heavily rely on constructor analysis due to the well-defined inheritance-revealing control flow during construction of a C++ object. 
	
	\vspace{.06in}
	\noindent
	{\em Optimization is common in C++ code, yet poses a serious impediment to class inheritance inference.}
	First, as a fundamental problem, constructor analysis suffers from low precision and is often insufficient.  
	This is because COTS C++ binaries are often optimized and tend to have many inlined functions including inlined constructors. 
	In fact, per ISO C++ 7.1.2/3---``{\em A function defined within a class definition is an inlined function}".
	Second, aggressive compiler optimization often results in exclusion of key functions (e.g., constructors) and/or entire classes from the binary, which makes inference hard. 
	For example, when a most derived class is not instantiated, the compiler may conveniently exclude such class definitions from the binary. 
	In fact, we consistently found a significant reduction in the number of constructors in the binary with higher levels of optimizations (see Table~\ref{tab:effect_inlining}). 
	Finally, it is hard to discern inherited relationship (e.g., class A inherits from class B) from composed relationship (e.g., class A contains an object of class B)--specially in the case of optimized code.
	
	These challenges are evidenced in most relevant recent works VCI~\cite{vci:2017} and Marx~\cite{marx:andre}.
	These efforts employ class hierarchy analysis on C++ binaries without relying on RTTI. 
	On the one hand, VCI's precision and accuracy are largely dependent on constructor identification, which in turn is heavily impeded by inlined or missing constructors. 
	On the other hand, Marx acknowledges the difficulty imposed by optimization and inlining, and limits its scope to identifying class membership to inheritance trees without actually recovering a directed class inheritance tree. 
	As a fundamental problem, aggressive compiler optimizations are common in COTS binaries, and pose complex challenges that state-of-the-art C++ binary analysis solutions are unable to handle. 
	
	In this paper, we present \codename, a robust class hierarchy inference engine for C++ binaries. 
	\codename\ employs static analysis and is built on top of BAP~\cite{Brumley:BAP}. 
	Unlike prior efforts, we support optimized code, which is common in COTS C++ programs. 
	As a key distinction, our inference engine is based on code features that {\em can not} be optimized away (i.e., eliminated) during compile time. 
	As such, these features form robust inference points. 
	\codename\ incorporates multiple novel analysis techniques in order to handle optimized code including inlined and missing constructors.
	This makes \codename\ apt for COTS binaries. 
	First, we take advantage of the fact that destructors tend to be virtual in COTS code as they help avoid memory leaks. 
	Because calls to virtual functions can not be statically resolved, compilers can not inline virtual functions during compile time, and retain them in the binary without inlining their code at the callsites. 
	Therefore, in comparison with constructors that are non-virtual, destructors in a binary tend to be more in number. 
	\codename\ employs a {\em constructor-destructor} combination approach to achieve optimal recovery.
	Second, because virtual functions must be retained in the binary, we employ inter-procedural object-layout analysis on virtual functions to construct an object model for each class and use it to recover inheritance relationship from functions with inlined constructors or inlined destructor, also eliminating false positives in inheritance relationship. We are the first to do this. 
	Finally, we identify precise points of completion of object construction in order to distinguish between composed and inherited objects. 
	To the best of our knowledge, this the first effort to effectively handle optimized C++ binaries commonly found in COTS software. 
	
\begin{figure}[ht]
		\centering
		\includegraphics[scale=0.2]{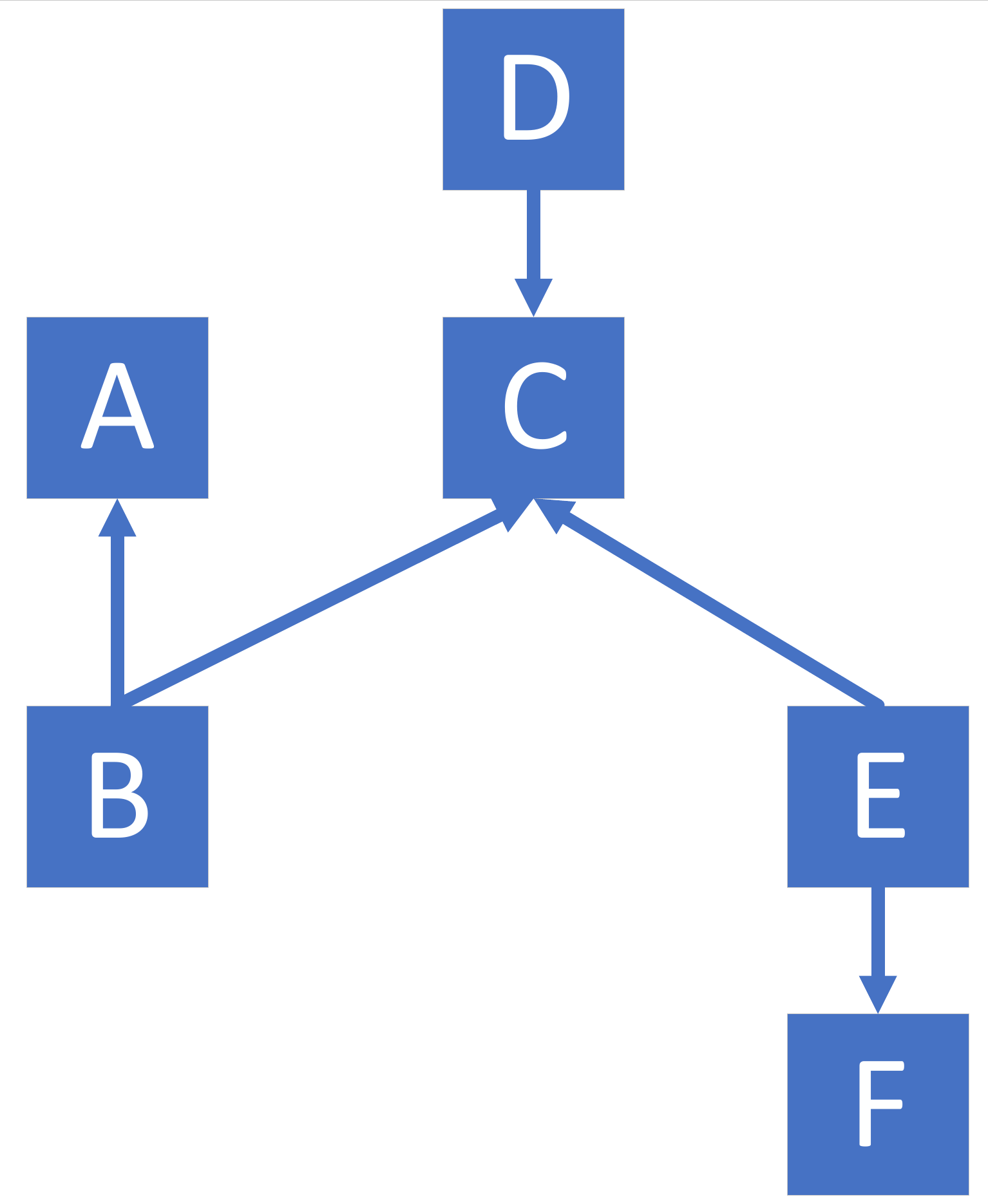}
		\caption{Class inheritance example illustrating the limitation of Marx.}\label{fig:marx_limitation}
	\end{figure}

	\vspace{.06in}
	\noindent
	{\bf State-of-the-art solutions are lacking.}
	The problem of class hierarchy recovery has been researched over the years, Marx and VCI being the more recent research solutions. While VCI identifies class relationships and direction of inheritance by strictly performing constructor analysis, Marx only identifies relationships by considering vptr writes and overwrites within an object. As we will explain more in depth in Section~\ref{sec:key_challenges}, VCI suffers from a high number of false positive and false negatives especially with optimized binaries(as reported in the paper). Marx is able to correctly identify related classes, however, it has a directionality problem. Marx groups unrelated classes together simply because there are one or more classes inheriting from them. Even when it groups only related classes, it still fails to reason about direction of inheritance. To illustrate the limitation of Marx, consider Figure \ref{fig:marx_limitation}. All the classes from A to F are grouped together as being related simply because there is a relationship between B and C and between E and F. Therefore, if Marx were to be used for CFI enforcement, for a callsite of static type A, an object with type of any of these 6 classes is considered valid (the source code for Marx is publicly available and we were able to verify this). This, to a large extent defeats the purpose of CFI.  \codename\ improves state-of-the-art solutions by performing constructor and destructor analysis to reduce false negatives, overwrite analysis to reduce false positives and false negatives as well as Object Layout Analysis(OLA) to assign direction of inheritance to relationships identified through overwrite analysis. With these techniques, \codename\ is able to correctly identify that B inherits only from A and C as well as the other inheritances.
	
	\vspace{.06in}
	\noindent
	{\bf Full CHT recovery is hard:}
	In general, recovering C++ semantics from optimized binaries is hard. Although \codename\ employs multiple novel techniques to handle optimized code, C++ compilers eliminate entire classes from the binary---if the classes are deemed to be unnecessary (e.g., through dead code elimination) during compilation.
	In such cases, \codename\ misses classes that have no remnants in the binary.
	Even so, to the best of our knowledge, \codename\ is the only {\em practical} solution that can effectively infer {\em directed} class inheritance tree from optimized code.
	
	Our contributions can be summarized as follows:
	\begin{itemize}
		\item We present \codename, an inference engine for recovering class hierarchy information from optimized C++ code. 
		\item We employ multiple novel analysis techniques including {\em constructor-destructor} analysis, inter-procedural object layout analysis, and precise identification of object completion. These techniques allow \codename\ to handle optimized code including inlined or missing constructors, distinguish between constructors and non-virtual destructors, and decipher between composed and inherited objects. The idea of extracting class hierarchy from both constructors and destructors have been considered previously ~\cite{fokin2010reconstruction}, however, for optimized code, we are the first to highlight the effectiveness of destructor analysis over constructor analysis.
		\item We evaluate \codename\ on 16 binaries. On average we are able to recover significant part of the inheritance graph correctly. Specifically, 94.5\% (recall) of edges under O0 and 71.4\% (recall) of edges under O2. 
	\end{itemize}
	
	The rest of the paper is organized as follows. In Section~\ref{sec:background}, we provide the technical background required to understand the rest of the paper. In Section~\ref{sec:overview} we present the key challenges and an overview of our solution. Section~\ref{sec:details} presents the technical details of our solution. We evaluate \codename\ in Section~\ref{sec:eval}, present related work in Section~\ref{sec:related}, and finally conclude in Section~\ref{sec:conclude}.

	\section{Technical Background}\label{sec:background}
	In this section, we provide the background relevant to rest of the paper.

	\begin{figure*}[ht]
		\centering
		\includegraphics[scale=0.5]{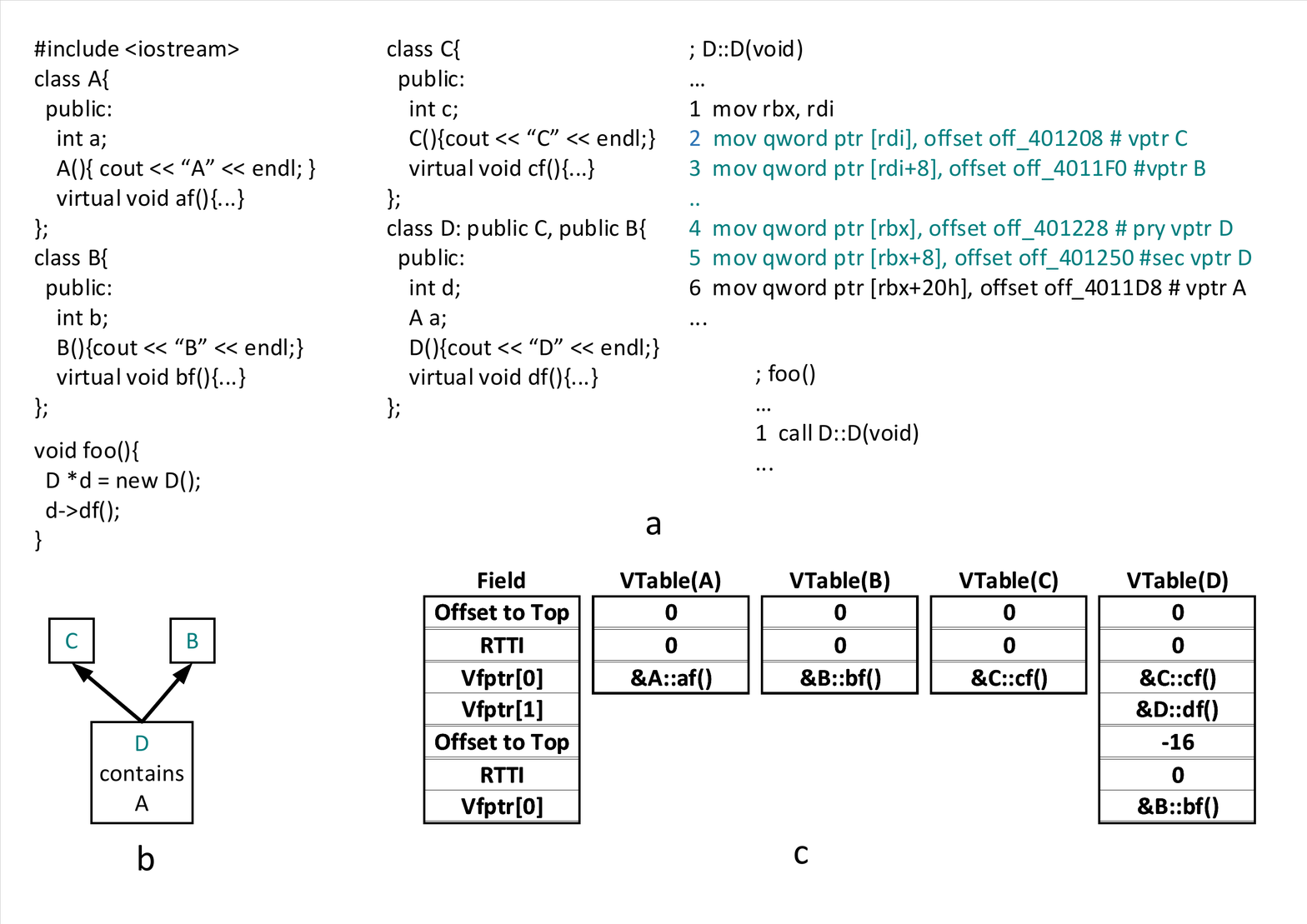}
		\caption{Running example comprising classes A, B, C and D, and corresponding VTable layout when compiled using g++ with flag {\tt -fno-rtti}. Note: `pry' means primary, and `sec' means secondary.}\label{fig:running}
	\end{figure*}
	
	\subsection{Implementation of Polymorphism and Inheritance in C++}
	In order to implement polymorphism, C++ compilers utilize a per-class supplementary data structure called ``VTable".
	The structure of the VTable is dictated by the C++ ABIs--- Itanium~\cite{ItaniumABI} and MSVC~\cite{Ray:1994:MSVC}.
	A VTable is allocated for each polymorphic class (i.e., a class that contains virtual functions, or inherits from class(es) that contain virtual function, or inherits a class virtually).
	Each VTable contains an array of function pointers representing the polymorphic functions that the object of a given type can invoke during runtime.
	In order to prevent corruption, the VTables are allocated in the read-only sections of a binary.
	Each object of a polymorphic class contains an implicit pointer to the VTable (vptr) as the first member variable. 
	Within the constructor of a polymorphic class, the vptr is initialized to point to the appropriate VTable for the type of the object being constructed. 
	Among other fields, each VTable contains 3 mandatory fields---{\em OffsetToTop}, pointer to \textit{TypeInfo} (also called RunTime Type Information or RTTI), and an array of virtual function pointers or vfptrs. 
	Although RTTI is a mandatory field, a NULL value is used to signify its exclusion during compilation. 
	In the past, mandatory fields have been used as a signature for identification of VTables in the binary~\cite{prakash:2015:vfguard}. 
	
	In case of multiple inheritance, wherein a class derives from more than one polymorphic base class, the VTable for derived class comprises of a group of 2 or more VTables---a primary and one or more secondary VTables depending on the number of secondary bases\footnote{The first base class in the declaration order is called the primary base, and the remaining bases are called secondary base classes} of the derived class.
	The derived class and its primary base class share the primary VTable, and each secondary base class is allocated a secondary VTable.
	Further, the derived object and the primary base sub-object share the same base address, and each secondary base sub-object is found at an offset from the derived object base address.
	The VTable group comprising of the primary and secondary VTables is collectively called ``complete-object VTable" for the derived class\footnote{VCI~\cite{vci:2017} uses the acronym VTT to refer to VTable group, which is fundamentally different from and inconsistent with the Virtual Table Tables defined in the Itanium ABI~\cite{ItaniumABI}. In this paper, we stick to the terminology used in the ABI.}. 
	
	The OffsetToTop field indicates the displacement that must be added to the sub-object within a derived object to reach the base of the derived object. 
	If the RTTI value is not null, the RTTI pointers for all the VTables within a complete-object VTable point to the same RTTI structure. 
	For the running example in Figure~\ref{fig:running}, the complete-object VTable for D comprises of a primary VTable with 2 virtual function pointers, and a secondary VTable with one virtual function pointer. 
	The secondary VTable contains an OffsetToTop value (-16) that must be added to the sub-object B-in-D to reach the base of D. 
	For more information on the need for each of the fields and other optional fields in the VTable, we refer readers to the ABI~\cite{ItaniumABI}. 
	
	Finally, by virtue of inheritance, C++ requires that derived class objects exhibit an {\em is-a} property with respect to each of the base class sub-objects.
	This is because, each member function in C++ accepts the object or {\tt this} pointer as an implicit argument and requires its type to be the same as class that defines the function. 
	The compiler embeds code to perform appropriate adjustments to objects (i.e., add/subtract offsets to reach to/from sub-objects to base object) to satisfy this property.
	
	\subsection{Construction and Destruction}\label{subsec:ctorAndDtor}
	Constructor and destructor are called when an object of a class is created and destroyed respectively. 
	Each class may contain one or more constructors and a single destructor.
	If no constructor is defined by the developer, a default constructor will be provided by the compiler. 
	The exact steps in the construction of an object are as follows:
	\begin{enumerate}
		\item[Step 1] Invoke constructors of bases starting from primary base and followed by each secondary base. The address of the subobject being constructed is passed as an argument. 
		\item[Step 2] Assign the vptr to object being constructed.
		\item[Step 3] Initialize the member variables including composed member objects, i.e., objects that are contained as members within the object being constructed. While the constructed object exhibits a {\em is-a} relationship with an inherited object, it exhibits a {\em has-a} relationship with a composed object. 
	\end{enumerate}
	Similarly, the exact steps in the destruction of an object are as follows:
	\begin{enumerate}
		\item[Step 1] Assign the vptr to the object being destructed.
		\item[Step 2] Destroy (i.e., finalize) the member variables including composed member objects. 
		\item[Step 3] Invoke destructors of immediate bases starting from primary base and each secondary base. The address of the subobject being destroyed is passed as an argument. 
	\end{enumerate}
	For non-polymorphic classes, the assignment of vptr is skipped. 
	As a part of the construction and destruction of member variables, constructors and destructors may themselves invoke virtual functions. 
	Therefore assignment of vptr occurs before the respective initialization/finalization of member variables. 
	Note that the constructors and destructors of each base class write their own set of vptrs in appropriate locations in the object, which get overwritten by subsequent classes in the inheritance chain.
	For example, in Figure~\ref{fig:running}, object D shares its base with subobject C-in-D (because C is the primary base of D).
	All the subobject constructors C() and B(), and composed object constructor A() are inlined in derived object constructor D().
	Instruction 2 writes vptr of C-in-D with address of VTable of C, which is then overwritten by instruction 4 that writes vptr of D with address of VTable of D.  
	Marx~\cite{marx:andre} performs ``overwrite-analysis" to leverage this behavior of constructors to construct groups of polymorphic types. 
	Compilers routinely inline constructors of base classes and constructors of composed member objects. 
	In Figure~\ref{fig:running}, instruction 6 initializes vptr of composed object A within D with address of VTable of A.  
	
	\subsection{Virtual Destructors}\label{subsec:virtdes}
	Unlike constructors, destructors in C++ can be--and often are--declared as virtual.
	It may sometimes be necessary to delete a derived class object that is referenced through a base class pointer. 
	The C++ standard states that deleting an object of derived class through a pointer to its base class that has non-virtual destructor leads to undefined behavior (see paragraph 3 in ISO/IEC 14882-2014).
	Therefore, it is common practice to mark destructors as virtual, which forces runtime resolution of the virtual call to the correct derived class destructor. 
	These destructors must therefore be retained in the binary and destructor code can not be inlined at the deletion site. 
	As we can see in Table~\ref{tab:effect_inlining}, the number of destructors in the binary are usually larger than the number of constructors.

	\section{Solution Overview}\label{sec:overview}

	\subsection{Key Challenges}\label{sec:key_challenges}
	Below, we enumerate key challenges in recovering high-level C++ semantics from an optimized binary.
	We use the running example in Figure~\ref{fig:running} to highlight the challenges. 
	
	\paragraph{C1: Constructor Inlining}
	Compilers inline functions by replacing the callsite with the body of the called function. Virtual function callsites cannot be inlined since the exact function to call is only known at runtime depending on the object type. However, because constructors can not be virtual, their calls are statically resolved and inlined when possible. 
	In fact, we found this to be very common and prescribed by the C++ standard (see ISO C++ 7.1.2/3). 
	Any function defined within a class definition will be inlined as a default behavior.
	This is a major challenge since state-of-the-art class hierarchy recovery tools like VCI~\cite{vci:2017} depend primarily on the identification of constructors and the operations they perform including invocation of base class constructors.
	
	Constructor inlining gives rise to two problems:
	\begin{itemize}
		\item {\em Missed base class constructor calls}: Consider the running example in Figure~\ref{fig:running} where the primary and secondary base class constructors of D are inlined within its constructor on lines 2 and 3 respectively. As detailed in section~\ref{subsec:ctorAndDtor}, in a constructor, the composed class' constructor get called after the vptrs of the owner class have been initialized. Therefore, in order not to include composed classes in a given class hierarchy, VCI looks at the first primary vptr initialization to an object address which appears on line 2 and concludes that the constructor belongs to the class with primary vptr 0x401208 (i.e., C instead of D), subsequent constructor calls are ignored. As such it fails to identify any relationship between D and C or D and B. Overwrite analysis adopted by Marx will be able to group the primary vptr of D with the primary vptr of C as well as the secondary vptr of D with the primary vptr of C, however, it cannot differentiate between the derived class and the base class.
		\item {\em False constructor identification}: In higher levels of optimization, the compiler could inline entire constructor D() in the instantiating function {\tt foo}. 
		Therefore, although not a constructor, {\tt foo} would contain the vptr initialization. 
		In order to accommodate inlining, VCI identifies constructors by simply looking for vptr initialization (not requiring that the vptr is written into the first entry of an object address) which would result in wrongly identifying {\tt foo} as a constructor (see Section 4.2 in ~\cite{vci:2017}). 
		If {\tt foo} calls other functions which also contain inlined constructors, a false relationship is inferred between the vptrs these non-constructor functions initialize. Table~\ref{tab:effect_inlining} shows how aggressive inlining occurs with high levels of optimization. As the optimization level increases, the number of constructors in the binary reduces.
		
	\end{itemize}
	
	\begin{table}[ht]
		\centering
		\caption{Table showing the number of constructors and destructors present in O0 and O2 binaries. It also shows functions with inlined constructors which other solutions could wrongly identify as constructors.}
		\label{tab:effect_inlining}
		\scalebox{0.9}{%
			\begin{tabular}{|l|l|l|l|l|l|l|}
				\hline
				\multicolumn{1}{|c|}{\multirow{2}{*}{Programs}} & \multicolumn{3}{c|}{O0} & \multicolumn{3}{c|}{O2} \\ \cline{2-7} 
				\multicolumn{1}{|c|}{} & Ctor & Dtor & \specialcell{Fns with\\ inlined \\ctor/dtor} & Ctor & Dtor & \specialcell{Fns with \\inlined\\ ctor/dtor} \\ \hline
				Mongodb & 3544 & 2063 & 40 & 725 & 1046 & 2638 \\
				Node & 2930 & 2546 & 75 & 290 & 447 & 2520 \\
				Doxygen & 1010 & 940 & 16 & 245 & 751 & 744 \\
				Soplex & 29 & 25 & 4 & 10 & 12 & 3 \\
				Povray & 47 & 24 & 9 & 40 & 19 & 20 \\
				Namd & 4 & 4 & 0 & 0 & 1 & 2 \\
				Omnetpp & 175 & 108 & 3 & 113 & 98 & 104 \\
				DealII & 582 & 702 & 48 & 390 & 677 & 497 \\
				Xalanc & 1391 & 958 & 309 & 954 & 771 & 1989 \\ \hline
				libebml & 41 & 18 & 26 & 42 & 18 & 25 \\
				libflac++ & 45 & 18 & 0 & 29 & 18 & 5 \\
				libzmq & 82 & 76 & 0 & 58 & 38 & 11 \\
				libwx\_baseu\_net & 47 & 44 & 2 & 22 & 35 & 33 \\
				libwx\_baseu & 371 & 287 & 0 & 158 & 257 & 209 \\
				libwx\_gtk2u\_adv & 310 & 259 & 12 & 48 & 155 & 240 \\
				libwx\_gtk2u\_aui & 74 & 59 & 0 & 18 & 46 & 55 \\
				libwx\_gtk2u\_core & 929 & 679 & 3 & 357 & 428 & 857 \\
				libwx\_gtk2u\_html & 140 & 136 & 0 & 26 & 66 & 91 \\
				libwx\_gtk2u\_xrc & 120 & 116 & 2 & 71 & 12 & 41 \\ \hline
			\end{tabular}
		}
	\end{table}
	
	\paragraph{C2: Inheritance vs Composition}
	A constructor (or destructor) not only calls the constructors (or destructor) of base classes, but also calls those of its composed classes (i.e., objects of classes contained as member variables).
	Failure to correctly differentiate the base class constructors (or destructors) from those of member classes will result in false inheritance inference between a class and its member classes. 
	VCI partially handles this by considering only constructor calls that happen before initialization of the primary vptrs, however, this works only for constructors but not destructors. As mentioned in C1, it does not guarantee base class vptr identification. In a destructor, VTable initialization happens first, followed by calls to composed class destructors and then to base class destructors with no demarcation between the two categories of destructor calls. A more general approach is required to demarcate composed and inherited objects for constructors or destructors irrespective of whether or not they are inlined. 
	
	\paragraph{C3: Missing Constructors}
	Another outcome of compiler optimization is complete removal of constructors. 
	Note that this is a different problem from function inlining. 
	A virtual function is guaranteed to be in the binary as long as the VTable it belongs to is present. This is not the case for non-virtual functions. 
	In fact, we found that significant number of constructors are optimized-out during compilation, and their definitions are excluded in the binary.
	For example, in Table~\ref{tab:effect_inlining} number of constructors present in binaries compiled with O2 is significantly less than number of constructors present in binaries compiled with O0.
	
	\subsection{Scope and Assumptions}
	Our primary goal is to leverage multiple binary-level features to reconstruct polymorphic non-virtual class inheritance tree from COTS binaries---even in an optimized setting. We target COTS C++ binaries which might have been compiled with high levels of optimization and with no debugging information, symbol information, RTTI, etc. Also, we assume that source code of such binaries are not available. Since our VTable extraction and construction call order are based on the ABI specification, we only target binaries compiled with standard C++ compilers, e.g., Clang, GCC and MSVC. 
	
	We aim to extract class inheritance only from benign binaries, therefore, we do not handle code obfuscation techniques which apart from constructor inlining could make certain information unavailable for analysis. 
	In addition, self-modifying or packed code or JIT-compiled code are out of the scope of this work. 
	
	\subsection{Our approach}
\begin{figure}[ht]
		\centering
		\includegraphics[scale=0.25]{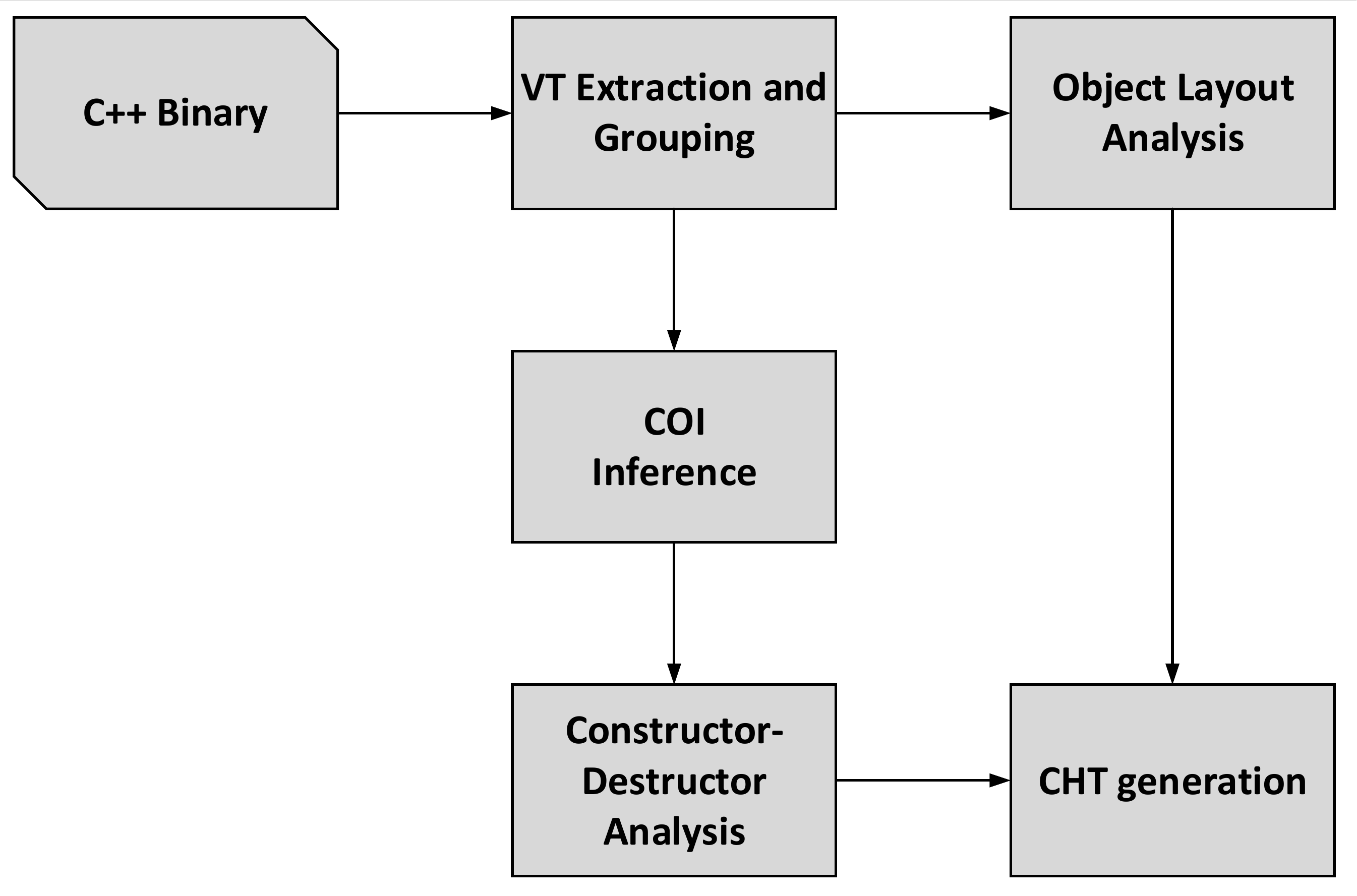}
		\caption{An overview of DeClassifier} \label{fig:overview}
	\end{figure}

	An overview of our approach is presented in Figure~\ref{fig:overview}. 
	As a preliminary step, we recover all the VTables in the binary and group them into complete-object VTables. 
	We utilize an already known scanning-based algorithm (used by vfGuard~\cite{prakash:2015:vfguard}, MARX~\cite{marx:andre}) to extract VTables. 
	These VTables include primary and secondary VTables, which are then grouped to form complete-object VTables. 
	In a nutshell, we start with a primary VTable (i.e., $offsetToTop == 0$), and group it with succeeding secondary VTables (i.e., $offsetToTop \ne 0$) until we reach the next primary VTable. 
	Each unique group is a complete-object VTable and provides a one-to-one mapping between the complete-object VTable and the polymorphic classes in the binary. In the remainder of this section, we outline the analyses we undertake to infer inheritance relationships between the complete-object VTables, i.e., polymorphic classes.
	
	\vspace{.06in}
	\noindent
	{\bf Key insight:} 
	Code features that require runtime decisions (e.g., virtual function dispatch) can not be optimized away (i.e., eliminated) by the compiler, and therefore provide a robust source for inheritance inference. 
	
	\vspace{.06in}
	\noindent
	Our inheritance inference approach is based on identifying key inference points that {\em can not} be optimized away during compilation (an exception being removal of entire classes).
	Particularly, we leverage the virtual functions including virtual destructors to infer inheritance semantics. 
	This way, we ensure meaningful inference even under strong compiler optimization. 
	
	\paragraph{Combining constructors with destructors}
	We combine constructor-destructor analysis to achieve optimal recovery. 
	State-of-the-art binary-level class inheritance extraction tools have primarily focused on constructor analysis. However, the number of constructors present in the binary decreases as the level of optimization increases, thus leading to inaccurate inference.  
	Like constructors, destructors also provide insight into a particular class inheritance. 
	Typically, the number of destructors in the binary tend to be larger than the number of constructors (see Table~\ref{tab:effect_inlining}).
	
	In order to prevent memory leakage, destructors are declared virtual which ensures that specific objects are destructed as expected. 
	
	\vspace{.06in}
	\noindent
	{\em Unlike constructors that can not be virtual, destructors are commonly virtual and therefore, destruction calls are not inlined during optimization.}
	
	\vspace{.06in}
	\noindent
	Because virtual functions are not inlined by compilers, explicit destructors are preserved in memory. 
	This eliminates the possibility of a virtual destructor not being present in the binary. 
	If we can also use destructors for our analysis, then we will be able to address challenge {\bf C1}, since we are no longer completely reliant on constructors--which could be inlined. 
	We could also address {\bf C3}, since we can augment destructors with available constructors.

	\paragraph{Identifying valid constructors and destructors}
	Inlining of constructors within other `host' functions results in false inference of the host function as a constructor. 
	This is one of the main reason for falses in analyzing optimized code. 
	In fact, vptrs initialized in one or more host functions that are neither constructors nor destructors will result in likely false relationship inference between such functions. 
	Therefore, if we can correctly eliminate such functions, we can safely analyze constructors and destructors, and correctly identify explicit calls to (or inlined) base constructors and destructors and also correctly eliminate calls to composed class constructors and destructors. This handles challenge {\bf C2}.
	Only the constructors assign vptr to the implicit {\em this} pointer.  
	We employ static analysis to detect whether the {\em this} pointer is initialized with a vptr.
	If so, it is classified as a constructor, if not it is a host function that contains inlined constructor.
	
	We also perform additional analysis, explained in Section~\ref{subsec:ctorDtorIdent} to differentiate constructors from destructors. 
	This helps us to address "false constructor identification" under {\bf C1}. 
	In order to address "missed base class constructor calls" for constructors which inline their base constructors, we ensure that the correct primary vptr associated to a constructor is identified. We do the same for destructors with inlined base destructors.
	
	\paragraph{Object Layout Analysis (OLA)}
	There are cases where destructors are not virtual, in that case, they could also be inlined just like constructors. This creates the possibility of some classes having neither constructor nor destructor present in the binary. 
	In cases where we can not find an explicit constructor or destructor for a class, we employ intra-procedural static analysis to model the object layout. 
	Specifically, we start from the explicit virtual functions in a class' VTable (note that these functions can not be optimized out).
	Next, we identify type-revealing instructions within the functions.
	Starting from these instructions, we obtain a backward slice and back-propagate type information to obtain a mapping:
	
	\vspace{-.02in}
	\begin{center}
		{\em this + offset -\textgreater type}
	\end{center}
	
	\vspace{-.02in}
	By extending the analysis across all the classes, and checking different class objects for type congruence (i.e., member types at a given offset across all polymorphic classes must be the same), we can eliminate inconsistent inferences.  
	
	We further our analysis by using information regarding pure virtual functions in the class VTables to improve recovery. Specifically, if VTable for class $A$ contains a pure virtual function at offset $off_1$, and VTable for class $B$ does not contain a pure virtual function at offset $off_1$, then A can not inherit from B.
	
	\paragraph{Identifying Completion of main Object Initialization (COI)}
	Completion of initialization of a class' main object (a.k.a main object initalization) is the point during construction where base class vptrs and class' vptrs have all been completely written and overwritten in the object. 
	Construction of composed objects always take place {\em after} construction of main object has completed. 
	Irrespective of compiler optimization, we can conclude that vptr initializations that occur before this point belong to base classes while those that occur after it belong to composed objects. 
	We found overwrite analysis~\cite{marx:andre} to be effective in identifying COI. 
	This saves us from relying on explicit base class constructor calls and also helps us to correctly eliminate composed objects from our inheritance. 
	In a case where a standalone constructor does not exist for a class, we depend only on the destructor. For classes with neither constructors nor destructors, we rely on the inference from OLA.
	
	In addition, completion of main object initialization makes it possible to correctly differentiate constructors and destructors from other functions which contain vptr initialization as a result of inlining. This subsequently helps to avoid false positive inheritance inference thereby solving challenge {\bf C1}.

	\section{DeClassifier}\label{sec:details}
	We have developed DeClassifier, a class inheritance inference engine that employs static analysis for reconstruction of class hierarchy from optimized C++ binaries. 
	It consists of 6 stages of analysis: VTable extraction and grouping, identification of completion of main object initialization, constructor-destructor analysis, object layout analysis, overwrite analysis and finally class hierarchy tree (CHT) generation.
	
	Below, we provide the technical details of each of the stages.
	
	\subsection{VTable Extraction and Grouping}\label{sec:vtable_grouping}
	As described in the background section, complete object VTables are made up of all the VTables that belong to a class, primary and secondary VTables. They provide an unlabeled unique representation for each polymorphic class in a given binary. Like other solutions which recover class hierarchy from binaries~\cite{marx:andre,vci:2017}, we treat complete object VTables as analogous to polymorphic classes and they form nodes in the CHT of a program generated by DeClassifier. 
	
	Much work has been done on extracting VTables from the binary~\cite{prakash:2015:vfguard,zhang:2015:vtint,gawlik:2014:tvip}. Vptrs are scattered throughout the text region of the binary as immediate values. Typically, they get written into locations in an object by constructors and destructors during object initialization.  So we scan the text section to recover all immediate values which point to read-only sections of memory, since VTables are stored in the read-only section to prevent VTable injection attacks. The well defined nature of VTables, particularly the existence of mandatory fields~\cite{ItaniumABI} provides us with a signature to filter out recovered immediate values which point to read-only section but are not VTables, for instance, jump tables.
	
	The recovered list of VTables contain all primary and secondary VTables where one or more of them make the complete object VTables for a single polymorphic class. Therefore, from the current list of VTables, we need to construct another list which comprises of only complete object VTables. To achieve this we merge primary VTables with their corresponding set of secondary VTables, with each item being represented by the primary VTable address. 
	
	All VTables belonging to a class are laid out contiguously starting from the primary VTable which has an offset-to-top of 0. All the secondary VTables have a non-zero offset-to-top. Given a set of VTables, we first sort them in increasing order of addresses. Then, we merge a primary VTable with all secondary VTables immediately following it. At the end of this process we have a set of complete object VTables each starting with a primary VTable followed by zero or more secondary Vtables.
	
	\subsection{\identificationOfCtorDtor}\label{subsec:ctorDtorIdent}
	Correctly differentiating constructors and destructors from other functions with inlined vptr initialization is crucial to eliminating all false positives. Constructor and destructor calls within actual constructors and destructors are those that guarantee inheritance. Functions containing inlined vptr initialization can contain multiple such initializations for different unrelated classes, therefore using any information within them will result in imprecise class hierarchy. 
	
	Constructors and destructors initialize the vptrs of the classes they belong to, among other operations they perform. The primary vptr of the class must be eventually written into the first entry of the object, before or after vptr of base classes are written, depending on whether a destructor or constructor is being considered. To do this, the object address gets passed to it, usually as the first argument. Therefore, we scan functions for primary vptr write to zero offset from the object address. 
	We lift the binary to BAP IR and construct use-def chains for each IR variable.
	Next, we recursively propagate the defines into uses until all IR instructions are a combination of defines corresponding to function inputs. 
	At this point, if the IR instruction corresponding to vptr initialization writes to the memory location pointed to by the first argument (implicit this pointer), we infer the function to be a constructor or a destructor. The instruction that writes vptr is the point of completion of object construction. 
	
	Our analysis can correctly distinguish constructors and destructors from functions which inline a constructor or destructor since the object address must be adjusted in order to write the primary vptr in the case of the latter. 
	This gives us the complete set of constructors and destructors. However, we are still left with the task of correctly differentiating between constructors and destructors so that we do not wrongly infer the derived class as the base or vice versa or include composed classes in the inheritance. As discussed in Section~\ref{subsec:ctorAndDtor}, the ordering of initialization of base and derived class vptrs are in the reverse order for constructors and destructors. 
	We infer a function to be a destructor if one of the following is true:
	
	\begin{enumerate}
		\item Destructors are mostly virtual, so they have entries in the VTables. We check if the function address exists in a VTable.
		\item Due to the use of destructors, for destructing objects, they call the delete operator. We also check if the function being verified calls the delete operator. A constructor will not call the delete operator.
		\item In cases where explicit calls are made to base class constructor, we check if the calls are made before vptrs are initialized.
	\end{enumerate}
	All identified constructors and destructors are associated with the primary vptr they belong to. In case of the constructor, the constructor is associated with the last primary vptr written to an object address. This is because if base class constructors are inlined, their vptrs are written first. For destructors, the destructor is associated with the first primary vptr written to the object address.
	Once we have the complete set of constructors and destructors, we move on to perform constructor-destructor analysis.
	
	\subsection{Constructor-Destructor Analysis}\label{subsec:ctorAndDtorAnalysis}
	The constructor of a derived class calls the constructors of its base classes (or initializes the base class vptrs) before initializing its own vptr(s). Since we have already identified valid constructors in the previous step, we extract all calls to valid constructors that take place before the last write of primary vptr to the first entry of an object. For inlined base class constructors as in the running example, we extract all complete object VTables (primary and secondary vptrs) initialized also before the last write of primary vptr. Composed classes get initialized, either through explicit constructor call or inlined vptr initialization, only after the complete object VTable of the current class has been initialized. Therefore, we are able to correctly exclude composed classes from our class hierarchy.
	
	In a destructor, the derived class' complete object VTable is first initialized, followed by calls to composed class destructors (or composed class vptr initialization) and finally, calls to base classes destructors. For a destructor, the last primary vptr write to zero offset of the object does not demarcate between base and composed classes. However, the number of secondary vptrs initialized gives us insight into where calls to composed class destructors end and where those of base classes destructors begin. The number of VTables a class has is equal to the total number of base classes (direct or indirect) it has. To correctly eliminate composed class destructors, we map each vptr initialized to each destructor call starting from the last call, since base class destructors are called last. Finally, we ignore other calls which do not have a corresponding vptr initialization. They are the composed class destructor calls which are in between the derived class vptr initialization and base class destructor calls.
	
	For the identified base class constructor/destructor calls, we locate their associated complete object VTable and map them as the base. In the case of inlined base class constructor/destructor, we directly map the inlined complete object VTable as the base.
	
	\begin{algorithm}[h]
		\caption{CtorAnalysis analyzes constructors in $\mathcal{C}$ to identify all base classes of each class whose constructor is analyzed.} \label{alg:ctorAnalysis}
		\begin{algorithmic}[1]
			\footnotesize
			\Procedure{CtorAnalysis}{$\mathcal{C}$}
			
			\For {{\bf each} $c$ {\bf in} $\mathcal{C}$}
			\State $ownerPryVT$ $\gets$ $getPrimaryVT(c)$
			\State $coi$ $\gets$ $getCOI(c)$
			\For {{\bf each} $instr$ {\bf in} $c$}
			\If {$isCall(instr) \&\& addressOf(instr) <= coi $}
			\State $target$ $\gets$ $getTarget(instr)$
			\If {$target in  \mathcal{C}$}
			\State $basePryVT \gets getPrimaryVT(target)$
			\State $Base{\{ownerPryVT\}}.append(basePryVT)$
			\EndIf
			\EndIf
			\EndFor
			\EndFor
			
			\Return $Base$
			
			\EndProcedure
			
		\end{algorithmic}
	\end{algorithm}
	
	\begin{algorithm}[h]
		\caption{DtorAnalysis analyzes destructors in $\mathcal{D}$ to identify all base classes of each class whose destructor is analyzed.} \label{alg:dtorAnalysis}
		\begin{algorithmic}[1]
			\footnotesize
			\Procedure{DtorAnalysis}{$\mathcal{D}$}
			\For {{\bf each} $d$ {\bf in} $\mathcal{D}$}
			\State $ownerPryVT$ $\gets$ $getPrimaryVT(d)$
			\State $coi$ $\gets$ $getCOI(d)$
			\State $noVTs \gets getNoOfVTs(d)$
			\For {{\bf each} $instr$ {\bf in} $d$}
			\If {$isCall(instr) \&\& addressOf(instr) <= coi $}
			\State $target$ $\gets$ $getTarget(instr)$
			\If {$target in  \mathcal{D}$}
			\State $allCalls.append(basePryVT)$
			\EndIf
			\EndIf
			\EndFor
			\For {$i = lenghtOf(allCalls) -1 \:to\: 0$}
			\State $basePryVT \gets getPrimaryVT(allCalls[i])$
			\State $Base_{ownerPryVT}.append(basePryVT)$
			\State $no_VTs = no_VTs - getNoOfVTs(allCalls[i])$
			\If {$!(noVTs > 0)$}
			\State {\bf break}
			\EndIf
			\EndFor
			\EndFor
			
			\Return $Base$
			
			\EndProcedure
			
		\end{algorithmic}
	\end{algorithm}
	
	\subsection{\objectAnalysis}\label{subsec:objectAnalysis}
	We perform object layout analysis on virtual member functions of a class.
	Particularly, we are interested in member functions that operate on the {\em this} pointer. 
	Virtual calls to these functions are explicit and can not be inlined, as such, they are available in the binary. 
	
	Specifically, we perform coarse type inferencing and label the object with its member types. 
	First, we convert the binary to BAP IR to perform static analysis. 
	Next, we identify type-revealing instructions in the function ({\tt jmp *ebx, mov rdi, rax; call printf,} etc.) and their corresponding IRs. 
	We employ intra-procedural static analysis to identify the offsets within {\em this} pointer that the types map to. 
	This approach is similar to the type inferencing performed by past efforts such as REWARDS~\cite{lin:2010:Rewards}.
	As an end result, we obtain a type map for offsets within {\em this} pointer. 
	In order for an inheritance relationship between two classes to be correct, types of member variables in the two classes at specific offsets must be congruent (compatible) to each other. 
	
	Next, we use overwrite analysis. For construction, base class vptrs are written followed by those of the derived class and the reverse is done for destruction. However, since there is no way to infer if an inlined vptr initialization belongs to a constructor or destructor, the order of overwrite cannot be used to infer direction of inheritance. Therefore, we use the result of OLA to decide direction of inheritance for relationships identified through overwrite analysis.
	We analyze specific attributes of an object as well as its complete object VTable. We consider presence of pure virtual functions, type congruence and VTable size.
	
	\paragraph{Pure Virtual Function}
	We perform pure virtual functions validation to further filter inaccuracies in inheritance inference.
	The presence or number of pure virtual functions in two VTables provides unidirectionality of inheritance since the VTable without pure virtual functions has to have derived from the VTable with pure virtual functions. These two cases are possible:

	\begin{enumerate}
		\item One VTable has pure virtual function entries while the other has none
		\item One VTable has more virtual function entries than the other, with both having those entries at the same offsets.
	\end{enumerate}
	A case where both VTables have the same number of pure virtual functions at the same offsets does not give any information about which of the two classes is the base. For this work, we consider both cases.

	\paragraph{Minimum Object Size Analysis}
	
	Analyzing the size of an object can be done either dynamically or statically. The dynamic analysis approach has two major challenges, 1) coverage and 2) how to compute size of stack and global objects. Objects are created in three major locations at runtime, heap, stack and global region of the memory. To create objects on the heap, the new operator must be called which can be hooked to get the size passed to malloc (that will be upper bound for the object size). However, size of stack objects pose a challenge in the sense that there is no difference between the stack pointer movement when memory is allocated for a local variable (e.g an integer variable) and for an object. 
	
	In this work, we analyze object size statically to obtain the lower bound of the size of polymorphic objects. With this approach, coverage is not a challenge and neither is the location of an object a challenge. Just like constructors and destructors, the first argument passed to member function of a class is the object address. To access a member variable, a literal value is added to the object address (i.e. \textit{this} pointer) to reach that variable. The maximum offset that can be added to the \textit{this} pointer will always be less than the size of the object itself, which gives a lower bound for the object size.  Identifying non virtual functions of a class is challenging, however, pointers to all virtual functions of a class reside in the complete object VTable for that class. For every complete object VTable identified, we analyze each virtual function it contains to obtain the maximum offset accessed from the \textit{this} pointer. We associate this value to the complete object VTable as the lower bound for its object size.
	
	\paragraph{VTable Offset to top}
	Considering the case of multiple inheritance, the derived object consists of sub-objects of its base classes. Offset-to-top refers to the offset of a given base-in-derived object from the top of the derived object. The offset-top-top value for each base class is stored in the offset-to-top field of the vtable corresponding to that base class. As already mentioned, one of the operations performed within a constructor is calling the constructor of base classes. Before this call is made, the constructor adds the offset-to-top value for the given base class to the \textit{thisptr} in order to reach the base class sub-object. Hence, we compare the offset-to-top value with the offset at the constructor call site for equality to conclude on inheritance between the two classes.
	
	\paragraph{VTable Size}
	We compute the VTable size of a class as the total number of virtual functions pointer, pure virtual functions and zero destructor entries (this exists only in the VTables of abstract classes) that it contain. We do not consider the complete object VTable in this case because, relationships can be identified between a secondary VTable and a primary VTable. Therefore we ensure that only the associated VTable sizes are considered.
	
	VTable sizes increase or remain steady down a particular inheritance chain, the VTable size of a derived class is always greater than or equal to that of its base class. Hence, the sizes of two VTables found to be related provide indication of direction of inheritance.
	
	\subsection{Performing overwrite analysis}
	We analyze each function identified to contain inlined constructor or destructor, examining every VTable writes (both primary and secondary) that they perform. VTable pointers written to the same memory locations are grouped together as being related. In Marx, if overwrite analysis identifies two vptrs A and B to be related and also finds B and C to be related, the three vptrs A, B and C are grouped to be in the same set even though A and C might not be related. In this work, once we identify a relationship between A and B, we immediately use the result from OLA to decide the direction of inheritance and then continue building up the class hierarchy with subsequent relationships. 
	
	We locate vptr overwrites in two ways, 1. if the object address passed to a known constructor or destructor is the same location where a primary or secondary vptr is written, and 2. if multiple vptrs are written in the same memory location. Note that these overwrites are considered on a function by function basis. For the first case, we locate the primary vptr associated with the constructor or destruction that is being called. 
	
	Object sizes are associated with the primary vptr of a complete object VTable. With multiple inheritance, a secondary vptr will overwrite a primary VTable or vice verse. Therefore, we also locate the corresponding primary VTable of every secondary VTable in any identified group of related vptrs. We are able to locate corresponding primary VTable with VTable grouping explained in Section~\ref{sec:vtable_grouping}. 
	
	\subsection{\cht Generation}\label{cht_gen}
	In this phase, we build the complete Class Hierarchy Tree for the binary being analyzed by combining the relationships identified during constructor-destructor analysis phase with those identified by overwrite analysis. Constructor-destructor analysis directly assigns direction to any relationship it recovers using the order of calls or vptr initialization. For relationships recovered through overwrite analysis, we use attributes obtained from OLA to assign direction of inheritance. 
	
	We apply the following rules to infer inheritance:
	\begin{itemize}
		\item If constructor of class A calls constructor B before completion of object construction, A inherits from B. Converse holds true for destructors. 
		\item Class A inherits from Class B only if size of object A $\ge$ size of object B.
		\item Class A inherits from Class B only if size of VTable of class A $\ge$ size of VTable of class B.
		\item Class A inherits from Class B only if type of each member in B given by $this_B + offset$ (from OLA) is compatible with the type of corresponding member $this_A + offset$ in A. 
		\item Class A inherits from Class B only if for each pure virtual function in VTable of A, the corresponding virtual function in VTable of B is also pure virtual.
		\item Class C is a secondary base of class A only if the {\tt offsetToTop} value in secondary VTable of A is equal to the displacement that must be added to an object of A to reach the C sub-object in A (obtained through OLA). For example, in the running example, the C sub-object in D can be obtained by adding 16 to the base address of D, which is nothing but the {\tt offsetToTop} value (-16) in secondary VTable of D.  
	\end{itemize}
	Once all decisions about direction of inheritance is made, we combine all edges to build the CHT of the binary.

	\section{Evaluation}\label{sec:eval}
	
	Our evaluation aims to answer the following questions:
	\begin{itemize}
		\item What fraction of the entire class hierarchy of a program can DeClassifier recover?
		\item How precise is the recovered class hierarchy?
		\item How effective is our direction of inheritance assignment using OLA?
		\item How does DeClassifier outperform constructor only analysis for recovering class hierarchy.
	\end{itemize}
	All binaries were compiled using gcc with O0 and O2 optimization. We do not consider O3 optimization for evaluation because in terms of inlining, similar binaries are produced with O2 and O3. All analysis experiments were performed on Ubuntu 16.04 LTS running on Intel core i7 3.60GHz with 32GB RAM. Experiments were performed on 10 libraries and 6 executables. We did not evaluate Node and Mongodb because BAP was unable to analyze them. Mongodb is too large for it and there was a runtime error while analyzing Node. We reported this error to BAP team and they acknowledged it is a bug in BAP which would be worked on. All SPEC2006 benchmark programs with polymorphic classes were considered except for Astar and Namd. Astar has just one polymorphic class, there is no edge for comparison. Namd has 3 polymorphic classes, where two of the classes inherit from the third. However, when compiled with O2 optimization, the VTable of the third gets optimized out.
	
	\begin{table*}[]
		\centering
		\caption{CHT recovery results for binaries compiled with gcc -O0. Column with ``inh = 0" contains \# of classes that do not inherit from any class, ``inh = 1" contains \# of classes that inherit from exactly 1 immediate base class, and ``inh \textgreater 1" contains \# of classes that inherit from more than 1 immediate base classes.}
		\label{opt_O0}
		\scalebox{0.9}{%
			\begin{tabular}{|l||llll||l|lll||lll|}
				\hline
				\multicolumn{1}{|c|}{\multirow{3}{*}{Programs}} & \multicolumn{4}{c|}{Ground Truth} & \multicolumn{7}{c|}{Analysis} \\ \cline{2-12} 
				\multicolumn{1}{|c|}{} & \multirow{2}{*}{\#Classes} & \multirow{2}{*}{inh=0} & \multirow{2}{*}{inh=1} & \multirow{2}{*}{inh\textgreater{}1} & \multirow{2}{*}{\#Classes} & \multicolumn{3}{c|}{Ctor only} & \multicolumn{3}{c|}{Ctor+Dtor}  \\ \cline{7-12} 
				\multicolumn{1}{|c|}{} &  &  &  &  &  & inh = 0 & inh = 1 & inh \textgreater 1 & inh = 0 & inh = 1 & inh \textgreater{}1  \\ \hline
				libebml & 27 & 5 & 22 & 0 & 26 & 14 & 12 & 0 & 7 & 19 & 0  \\
				libflac & 18 & 8 & 10 & 0 & 18 & 12 & 6 & 0 & 8 & 10 & 0  \\
				libzmq & 76 & 17 & 47 & 12 & 76 & 39 & 29 & 8 & 17 & 47 & 12  \\
				libwx\_baseu & 285 & 24 & 258 & 3 & 287 & 157 & 128 & 2 & 49 & 235 & 3  \\
				libwx\_baseu\_net & 44 & 27 & 15 & 2 & 44 & 35 & 8 & 1 & 27 & 15 & 2 \\
				libwx\_gtk2u\_adv & 266 & 150 & 114 & 2 & 266 & 199 & 67 & 0 & 146 & 117 & 3  \\
				libwx\_gtk2u\_aui & 62 & 51 & 11 & 0 & 62 & 58 & 4 & 0 & 52 & 9 & 1  \\
				libwx\_gtk2\_core & 683 & 220 & 445 & 18 & 683 & 408 & 263 & 12 & 242 & 419 & 22  \\
				libwx\_gtk2u\_html & 138 & 66 & 70 & 2 & 138 & 104 & 34 & 0 & 65 & 71 & 2  \\
				libwx\_gtk2u\_xrc & 122 & 110 & 12 & 0 & 122 & 116 & 6 & 0 & 111 & 11 & 0  \\ \hline
				Doxygen & 974 & 208 & 670 & 96 & 944 & 462 & 417 & 65 & 222 & 629 & 93  \\
				Xalanc & 975 & 317 & 643 & 15 & 968 & 472 & 486 & 10 & 308 & 646 & 14  \\
				DealII & 884 & 34 & 846 & 4 & 689 & 431 & 256 & 2 & 47 & 639 & 3  \\
				Omnetpp & 112 & 10 & 102 & 0 & 109 & 50 & 59 & 0 & 11 & 98 & 0  \\
				Soplex & 29 & 8 & 20 & 1 & 29 & 18 & 11 & 0 & 8 & 20 & 1  \\
				Povray & 32 & 11 & 21 & 0 & 29 & 16 & 13 & 0 & 12 & 17 & 0  \\ \hline
			\end{tabular}
		}
	\end{table*}
	
	\begin{table*}[]
		\centering
		\caption{CHT recovery results for binaries compiled with gcc -O2. Column with ``inh = 0" contains \# of classes that do not inherit from any class, ``inh = 1" contains \# of classes that inherit from exactly 1 immediate base class, and ``inh \textgreater 1" contains \# of classes that inherit from more than 1 immediate base classes.}
		\label{opt_O2}
		\begin{tabular}{|l||l||lll||lll||lll|}
			\hline
			\multirow{2}{*}{Programs} & \multirow{2}{*}{\#Classes} & \multicolumn{3}{l|}{Ctor only} & \multicolumn{3}{l|}{Ctor+Dtor} & \multicolumn{3}{l|}{\specialcell{Ctor+Dtor +\\Object Layout Analysis (OLA)}} \\ \cline{3-11} 
			&  & inh = 0 & inh = 1 & inh \textgreater 1 & inh = 0 & inh = 1 & inher \textgreater{}1 & inh = 0 & inh = 1 & inh \textgreater{}1 \\ \hline
			libebml & 26 & 14 & 12 & 0 & 7 & 19 & 0 & 7 & 19 & 0 \\
			libflac & 18 & 15 & 3 & 0 & 8 & 10 & 0 & 8 & 10 & 0 \\
			libzmq & 63 & 35 & 24 & 4 & 23 & 37 & 4 & 22 & 40 & 2 \\
			libwx\_baseu & 262 & 233 & 29 & 0 & 171 & 91 & 0 & 153 & 108 & 1 \\
			libwx\_baseu\_net & 43 & 37 & 6 & 0 & 29 & 14 & 0 & 29 & 14 & 0 \\
			libwx\_gtk2u\_adv & 229 & 214 & 15 & 0 & 197 & 18 & 0 & 193 & 35 & 1 \\
			libwx\_gtk2u\_aui & 59 & 57 & 2 & 0 & 57 & 2 & 0 & 54 & 5 & 0 \\
			libwx\_gtk2\_core & 621 & 566 & 55 & 0 & 486 & 135 & 0 & 364 & 240 & 17 \\
			libwx\_gtk2u\_html & 123 & 118 & 5 & 0 & 104 & 19 & 0 & 104 & 19 & 0 \\
			libwx\_gtk2u\_xrc & 93 & 93 & 0 & 0 & 93 & 0 & 0 & 93 & 0 & 0 \\ \hline
			Doxygen & 870 & 845 & 23 & 0 & 458 & 400 & 12 & 483 & 446 & 5 \\
			Xalanc & 875 & 615 & 258 & 2 & 396 & 479 & 0 & 410 & 457 & 8 \\
			DealII & 687 & 544 & 140 & 3 & 123 & 561 & 3 & 96 & 579 & 11 \\
			Omnetpp & 105 & 69 & 36 & 0 & 36 & 69 & 0 & 8 & 91 & 5 \\
			Soplex & 25 & 23 & 2 & 0 & 21 & 3 & 1 & 19 & 6 & 0 \\
			Povray & 24 & 21 & 3 & 0 & 17 & 7 & 0 & 17 & 7 & 0 \\ \hline
		\end{tabular}
	\end{table*}

	\begin{table*}[ht]
		\centering
		\caption{Precision and Recall of CHT generated by \codename\ for O0 Optimization with GCC. Comparison was done with the overall ground truth}
		\label{percent_O0}
		\begin{tabular}{|l|l|l|ll|ll|}
			\hline
			\multirow{2}{*}{Programs} & \multirow{2}{*}{\specialcell{\#Classes\\in GT}} & \multirow{2}{*}{\specialcell{\#Classes\\in Binary}} & \multicolumn{2}{l|}{Ctor only} & \multicolumn{2}{l|}{Ctor+Dtor} \\ \cline{4-7} 
			&  &  & Precision(\%) & Recall(\%) & Precision(\%) & Recall(\%)  \\ \hline
			libebml & 27 & 26 & 100 & 54.5  & 100 & 86.4  \\
			libflac & 18 & 18 & 100 & 60  & 100 & 100  \\
			libzmq & 76 & 76 & 95.7 & 59.2  & 97.3 & 96.1  \\
			libwx\_baseu & 285 & 287 & 99.2 & 49 & 98.3 & 89.8  \\
			libwx\_baseu\_net & 44 & 44 & 100 & 52.6  & 100 & 100  \\
			libwx\_gtk2u\_adv & 266 & 266 & 94.0 & 50  & 91.9  & 95.8  \\
			libwx\_gtk2u\_aui & 62 & 62 & 100 & 44.4  & 90.9 & 90.9  \\
			libwx\_gtk2\_core & 683 & 683 & 98.6 & 58 &97.2  & 93.8  \\
			libwx\_gtk2u\_html & 138 & 138 &100  & 48.6  & 98.7 & 100  \\
			libwx\_gtk2u\_xrc & 122 & 122 & 100 & 45.5  & 100 & 91.7  \\ \hline
			\textbf{Average} &  &  & \textbf{98.8} & \textbf{52.2}  & \textbf{97.4} & \textbf{94.5}  \\ \hline
			Doxygen & 974 & 944 & 99.8 & 63.4  &98.9  & 93.5  \\
			Xalanc & 975 & 968 & 100 & 70.4  & 100 & 97.5  \\
			DealII & 874 & 689 & 95 & 28.9  &99.8  & 75.3  \\
			Omnetpp & 112 & 109 & 100 & 57.8 & 100 & 96.1  \\
			Soplex & 29 & 29 & 100 & 50 & 100  & 100  \\
			Povray & 32 & 29 & 100 & 61.9  & 100 & 81.0  \\ \hline
			\textbf{Average} &  &  & \textbf{99.13} & \textbf{55.4}  & \textbf{99.8} & \textbf{90.6}  \\ \hline
		\end{tabular}
	\end{table*}
	
	\begin{table*}[]
		\centering
		\caption{Precision and Recall of CHT generated by \codename\ for O2 Optimization with GCC. Comparison was done with those edges in ground truth whose corresponding VTables were found in the binary}
		\label{percent_O2}
		\begin{tabular}{|l|ll|ll|ll|ll|ll|}
			\hline
			\multirow{2}{*}{Program} & \multicolumn{2}{l|}{\#Classes} & \multicolumn{2}{l|}{\#Edges} & \multicolumn{2}{l|}{Ctor only} & \multicolumn{2}{l|}{Ctor + Dtor} & \multicolumn{2}{l|}{Ctor + Dtor + OLA} \\ \cline{2-11} 
			& GT & Binary & GT & Used & Precision(\%) & Recall(\%) & Precision(\%) & Recall(\%) & Precision(\%) & Recall(\%) \\ \hline
			libebml & 27 & 26 & 22 & 22 & 100.0 & 54.6 & 100.0 & 86.4 & 100.0 & 86.4 \\
			libflac & 18 & 18 & 10 & 10 & 100.0 & 30.0 & 100.0 & 100.0 & 100.0 & 100.0 \\
			libzmq & 76 & 64 & 76 & 53 & 100.0 & 60.4 & 97.8 & 75.5 & 100.0 & 79.3 \\
			libwx\_baseu & 285 & 262 & 264 & 198 & 100.0 & 14.1 & 100.0 & 43.9 & 100.0 & 47.5 \\
			libwx\_baseu\_net & 44 & 43 & 19 & 17 & 100.0 & 35.3 & 92.9 & 76.5 & 100.0 & 82.4 \\
			libwx\_gtk2u\_adv & 266 & 229 & 118 & 83 & 100.0 & 18.1 & 88.2 & 18.1 & 91.4 & 38.6 \\
			libwx\_gtk2u\_aui & 62 & 59 & 11 & 11 & 50.0 & 11.1 & 50.0 & 11.1 & 80.0 & 44.4 \\
			libwx\_gtk2\_core & 683 & 621 & 481 & 293 & 95.1 & 13.3 & 94.7 & 30.4 & 93.8 & 61.4 \\
			libwx\_gtk2u\_html & 138 & 123 & 74 & 36 & 100.0 & 13.9 & 88.9 & 44.4 & 89.5 & 47.2 \\
			libwx\_gtk2u\_xrc & 122 & 102 & 12 & - & 0.0 & 0.0 & 0.0 & 0.0 &0.0  & 0.0 \\ \hline
			\textbf{Average} &  &  &  &  & \textbf{84.5} & \textbf{25.1} & \textbf{81.3} & \textbf{48.6} & \textbf{85.4} & \textbf{58.4} \\ \hline
			Doxygen & 974 & 870 & 866 & 469 & 100.0 & 3.0 & 68.2 & 57.6 & 94.7 & 80.2 \\
			Xalanc & 975 & 875 & 710 & 577 & 100.0 & 45.4 & 78.7 & 65.3 & 98.3 & 79.4 \\
			DealII & 874 & 687 & 854 & 678 & 98.4 & 18.6 & 99.1 & 80.1 & 98.4 &81.9  \\
			Omnetpp & 112 & 105 & 102 & 97 & 100.0 & 22.7 & 100.0 & 58.8 & 98.7 & 78.4 \\
			Soplex & 29 & 25 & 22 & 12 & 100.0 & 8.3 & 66.7 & 16.7 & 100.0 & 50.0 \\
			Povray & 32 & 24 & 21 & 12 & 100.0 & 23.1 & 100.0 & 58.3 & 100.0 & 58.3 \\ \hline
			\textbf{Average} &  &  &  &  & \textbf{99.7} & \textbf{20.2} & \textbf{85.5} & \textbf{56.1} & \textbf{98.4} & \textbf{71.4} \\ \hline
		\end{tabular}
	\end{table*}
	\begin{table}[]
		\centering
		\caption{Number of direction of inheritance ``correctly assigned", ``wrongly assigned" and not ``assigned" by OLA}
		\label{ola_accuracy}
		\begin{tabular}{|l|l|l|l|}
			\hline
			Programs & Correctly assigned & Wrongly assigned & Not assigned \\ \hline
			libebml & 19 & 0 & 0 \\
			libflac & 10 & 0 & 0 \\
			libzmq & 42 & 0 & 2 \\
			libwx\_baseu & 94 & 0 & 0 \\
			libwx\_baseu\_net & 14 & 0 & 0 \\
			libwx\_gtk2u\_adv & 32 & 0 & 8 \\
			libwx\_gtk2u\_aui & 4 & 0 & 0 \\
			libwx\_gtk2\_core & 171 & 2 & 20 \\
			libwx\_gtk2u\_html & 17 & 0 & 1 \\ \hline
			Doxygen & 373 & 2 & 49 \\
			Xalanc & 459 & 0 & 17 \\
			DealII & 555 & 11 & 10 \\
			Omnetpp & 73 & 4 & 0 \\
			Soplex & 6 & 0 & 1 \\
			Povray & 7 & 2 & 0 \\ \hline
		\end{tabular}
	\end{table}
	
	\subsection{Ground Truth}
	We obtained the ground truth for standalone programs by compiling them with the {\tt -fdump-class-hierarchy} option on GCC. This generates a .class file for each .cpp file with at least one polymorphic class which contains VTables as well as the inheritance of those classes. However, since the 7 WX Widget programs in our test set are together in a single package, there is no way we could distinguish the classes that belong to each of the program using the output of {\tt -fdump-class-hierarchy}. Therefore, we compiled the program with the {\tt -frtti} option with no optimization and then analyzed the RTTI structures in each of the binaries to obtain the ground truth inheritance.
	\subsection{Precision and Recall}
	In order to measure the performance of \codename\, we evaluated precision \textit{P} and Recall \textit{R} of the class hierarchy recovered from each of the 16 binaries considered. Precision answers the question of what fraction of the class hierarchy recovered is correct and what fraction is wrong. We defined it as follows:
	\begin{equation}
	P=\frac{TP}{TP+FP}
	\end{equation}
	while recall answers the question of what fraction of the ground truth class hierarchy has been recovered and what fraction is not recovered. It is defined as follows:
	\begin{equation}
	R=\frac{TP}{TP+FN}
	\end{equation}
	where \textit{TP}, \textit{FP} and \textit{FN} refer to the number of derived-to-base edges recovered which match edges in the ground truth, number of edges which do not match any in the ground truth and number of edges in the ground truth not recovered.
	Table~\ref{opt_O0} and Table~\ref{opt_O2} show the breakdown of classes based on the number of classes they inherit from for O0 and O2 optimization levels respectively. Table~\ref{percent_O0} and Table~\ref{percent_O2} show the precision and recall of Ctor only analysis, Ctor+Dtor analysis as well as Ctor+Dtor+OLA for O0 and O2 optmization levels respectively.
	
	Recall for O0 binaries was computed using all found edges in the ground truth. However, this is done differently for O2 binaries.  Due to optimization, some classes (VTables) get removed by the compiler and the fact that our tool does not recover such classes does not make it less effective since they are in fact not available in the binary. To ensure these classes do not influence the recall recorded, we identified them and removed edges which have them as either derived or base from the ground truth to compare with. Basically, for O2 binaries, we computed recall by comparing with a subset ground truth which is based on the available classes(VTable) found in the binary. Columns labeled "GT" and "Used" under "\#Edges" in Table~\ref{percent_O2} show the number of edges in the overall ground truth and the number of edges remaining after removing edges with classes not found in the binary.
	
	The average precision and recall of class hierarchy recovered by \codename\ on O0 binaries are 97.4\% and 94.5\% for libraries and 99.8\% and 90.6\% for executables respectively. And on O2 binaries, it has an average precision and recall of 85.4\% and 58.4\% for libraries and 98.4\% and 71.4\% for executables  respectively. \codename\ was unable to recover any hierarchy for libwx\_gtk2u\_xrc.

	\subsection{Effectiveness of Direction Assignment using OLA}
	In this subsection, we discuss the effectiveness of direction of inheritance assignment using OLA. As discussed in section~\ref{cht_gen}, after identifying a relationship between two classes using overwrite analysis, we use OLA to assign the appropriate direction of inheritance. Table~\ref{ola_accuracy} shows the number of directions correctly assigned, the number wrongly assigned (i.e. assigning the derived as the base) and the number not assigned at all. We do not assign direction of inheritance between two classes whenever there is no enough information from OLA about those classes. On the average, directions of inheritance were correctly assigned to  93.6\% of relationships identified, 1\% were wrongly assigned and 5.4\% were not assigned at all.

	\subsection{Comparison with Constructor only Analysis} 
	Table~\ref{percent_O0} and Table~\ref{percent_O2} show how recall significantly increases from Ctor only analysis, to Ctor-Dtor + OLA. Precision decreases slightly from Ctor only to Ctor-Dtor, but increases again for Ctor-Dtor + OLA analysis. Such decrease in precision is recorded because, for destructor analysis, vptrs are mapped to base class destructors starting from the last call. As a result, if a class has no base class, but has a template class, the template class will be wrongly identified as its base class. However, with overwrite analysis, we are able to see that no overwrite actually happens between the two vptrs involved. First, the combination of destructor and constructor significantly increased the recovery compared to using constructors alone. Secondly, combining these with OLA helped to recover all other details in the binary that are unavailable from either constructor or destructor analysis as a result of optimization. For O0 binaries, the average recall increased from 52.2\% to 94.5\% for libraries and from 55.4\% to 90.6\% for executables for Ctor only and Ctor-Dtor analysis respectively. Since no optimization is performed on O0 binaries, overwrite analysis improved neither precision nor recall, for this reason, we did not include a different column for overwrite analysis. For O2 binaries, the tables show that recall increased from 25.1\% to 48.6\% to 58.4\% for libraries and from 20.2\% to 56.1\% to 71.4\% for executables for Ctor only, Ctor-Dtor and Ctor-Dtor + OLA respectively.
	
	\subsection{Comparison with other Class Hierarchy Recovery Solutions}
	Table~\ref{tab:related} summarizes existing class hierarchy recovery solutions based on their capabilities and techniques. The following were considered:
	\begin{enumerate}
		\item HandleInlining: Ability to correctly identify relationships when constructors or destructors are inlined either in the constructors or destructors of their derived classes or in other functions.
		\item InhVsComp: Ability to correctly differentiate inherited objects from composed objects.
		\item DtorAnalysis: The use of destructor analysis in class hierarchy recovery.
		\item CHTRecovery: The format in which class hierarchy tree is recovered. The edge from a class to another class can either be directed indicating that one inherits from the other, or undirected. All related classes can also be grouped together or not.
	\end{enumerate}
	
	We were able to run Marx on a number of binaries. We found out that it can handle 1 and 2 however, the class hierarchy recovered is undirected where related classes are grouped into sub trees. The approach proposed by Katz et al\cite{katz:2016} is similar to Marx based on the features we consider here. We contacted the authors of VCI for the source code but they did not release it to us. From the description given in the literatures for VCI, SmartDec and Hex Rays, we see that they are unable to handle 1 and they can correctly handle 2 only when there is no optimization. Also, they do not consider 3 and all hierarchy recovered are directed. ObjDigger does not handle any of the capabilities we considered. Lego handles inlined constructors, but does not handle 2 and does not consider 3. Recovered CHT are directed.
	
	\begin{table*}[ht]
		\centering
		\caption{Solutions that extract \underline{at least partial} Class Hierarchy Tree (CHT) from C++ binaries, their key techniques and limitations. Entry $p$ implies recovery only when ctors not inlined.}
		\label{tab:related}
		\begin{tabular}{|l|l|c|c|c|l|}
			\hline
			{\bf Solution} & \specialcell{{\bf Key }{\bf Technique}} & \specialcell{{\bf HandleInlining}} & \specialcell{{\bf InhVsComp}} & \specialcell{{\bf DtorAnalysis}}&\specialcell{{\bf CHTRecovery}} \\ \hline
			VCI\cite{vci:2017} & \specialcell{Static ctor analysis} & \xmark & $p$ & \xmark & Directed\\ \hline
			Marx\cite{marx:andre} & \specialcell{Static overwrite analysis} & \cmark & \cmark & \xmark & \specialcell{Undirected\\Sub-Tree Grouping}\\ \hline
			Katz et al.\cite{katz:2016} & \specialcell{Object tracelet, predictive modeling} & \cmark & \cmark & \xmark & \specialcell{Undirected\\Sub-Tree Grouping}\\ \hline
			SmartDec\cite{fokin:2011:C++Rev,fokin2010reconstruction} & \specialcell{Static ctor analysis} & \xmark & $p$ & \xmark & Directed \\ \hline
			ObjDigger\cite{Jin:2014:objdigger} &Symbolic execution, data flow analysis & \xmark & \xmark & \xmark & Not applicable \\ \hline
			
			Lego\cite{srinivasan2013software} & \specialcell{Dynamic overwrite analysis} & \cmark & \xmark & \xmark& Directed \\ \hline
			\specialcell{Hex Rays\cite{igor:ida:decompiler}} & \specialcell{Static ctor analysis} & \xmark & $p$ & \xmark & Directed \\ \hline
			{\bf DeClassifier} & Static ctor-dtor, overwrite analysis & \cmark & \cmark & \cmark & Directed \\ \hline
		\end{tabular}
	\end{table*}
	
	\begin{table*}[]
		\centering
		\caption{Number of false positives recovered by Ctor only analysis, Ctor-Dtor analysis and Ctor-Dtor + OLA analysis for O2 binaries. ``Total edges not found" refers to edges in \codename's output but not in ground truth. ``MIB" refers to Missing Immediate Base. ``Actual false positive" refers to actual false positive edges recovered by \codename}
		\label{false_positives}
		\begin{tabular}{|l|lll|lll|lll|}
			\hline
			\multirow{2}{*}{Programs} & \multicolumn{3}{c|}{Ctor only} & \multicolumn{3}{c|}{Ctor + Dtor} & \multicolumn{3}{c|}{Ctor + Dtor + OLA} \\ \cline{2-10} 
			& \specialcell{Total edges\\ not found} & \specialcell{Edges \\from MIB} & \specialcell{Actual False\\ positive} & \specialcell{Total edges\\ not found} & \specialcell{Edges \\from MIB} & \specialcell{Actual False\\ positive} & \specialcell{Total edges \\not found} & \specialcell{Edges \\from MIB} & \specialcell{Actual False\\ positive} \\ \hline
			libebml & 0 & 0 & 0 & 0 & 0 & 0 & 0 & 0 & 0 \\
			libflac & 0 & 0 & 0 & 0 & 0 & 0 & 0 & 0 & 0 \\
			libzmq & 0 & 0 & 0 & 1 & 0 & 1 & 0 & 0 & 0 \\
			libwx\_baseu & 1 & 1 & 0 & 4 & 4 & 0 & 15 & 15 & 0 \\
			libwx\_baseu\_net & 0 & 0 & 0 & 1 & 0 & 1 & 0 & 0 & 0 \\
			libwx\_gtk2u\_adv & 0 & 0 & 0 & 3 & 1 & 2 & 4 & 1 & 3 \\
			libwx\_gtk2u\_aui & 1 & 0 & 1 & 1 & 0 & 1 & 1 & 0 & 1 \\
			libwx\_gtk2\_core & 16 & 14 & 2 & 48 & 43 & 5 & 90 & 76 & 14 \\
			libwx\_gtk2u\_html & 0 & 0 & 0 & 3 & 1 & 2 & 2 & 0 & 2 \\
			libwx\_gtk2u\_xrc & 0 & 0 & 0 & 0 & 0 & 0 & 0 & 0 & 0 \\ \hline
			Doxygen & 9 & 9 & 0 & 142 & 16 & 126 & 77 & 53 & 24 \\
			Xalanc & 0 & 0 & 0 & 102 & 0 & 102 & 6 & 1 & 5 \\
			DealII & 20 & 18 & 2 & 24 & 19 & 5 & 52 & 32 & 20 \\
			Omnetpp & 14 & 14 & 0 & 12 & 12 & 0 & 24 & 21 & 3 \\
			Soplex & 1 & 1 & 0 & 3 & 2 & 1 & 0 & 0 & 0 \\
			Povray & 0 & 0 & 0 & 0 & 0 & 0 & 0 & 0 & 0 \\ \hline
		\end{tabular}
	\end{table*}

	\section{Discussion}\label{sec:discuss}
	\subsection{Falses---Root Cause Analysis}
	\begin{table}[ht]
		\centering
		\caption{Column 2 shows the total number of classes that make up the missing edges for O2 optimization, column 3 shows the number of those classes whose VTables were not found in the binary}
		\label{missing_bases}
		\begin{tabular}{|l|l|l|}
			\hline
			Programs & \specialcell{\#Polymorphic\\Classes\\missing edges ({\bf Col A})} & \specialcell{\#Classes in {\bf Col A}\\without VTables in Bin} \\ \hline
			libzmq & 20 & 13 \\
			libwx\_baseu & 88 & 27 \\
			libwx\_baseu\_net & 6 & 3 \\
			libwx\_gtk2u\_adv & 63 & 22 \\
			libwx\_gtk2u\_aui & 5 & 5 \\
			libwx\_gtk2u\_core & 198 & 68 \\
			libwx\_gtk2u\_html & 13 & 7 \\ \hline
			Doxygen & 178 & 100 \\
			Xalanc & 158 & 84 \\
			Omnetpp & 13 & 5 \\
			DealII & 257 & 169 \\
			Soplex & 3 & 2 \\
			Povray & 16 & 4 \\ \hline
		\end{tabular}
	\end{table}
	
	\paragraph{Inaccuracies in static disassembly}
	We did not expect to find any false relationship from overwrite analysis, however, this is as a result of inaccuracies in static disassembly.
	In the disassembly produced by BAP for the snippet below, the sequence of instructions after 4 is 6, 7, 5. But there is no actual branch instruction to go back to 5 if the jump is executed. At 7, the content of rbx is stored back in rdi, which is the same location where 0xEDFDA8 is stored at 2. As a result, the rdi value passed to the destructor of InheritedMemberInfoContext::Private is the same as the location where 0xEDFDA8 is written. Therefore our overwrite analysis identifies 0xEDFDA8 and the primary vptr of InheritedMemberInfoContext::Private as being related whereas they are not. If this were dynamic analysis, instruction 5 will not be executed if the jump is executed. 
	
\begin{footnotesize}
\begin{lstlisting}
 ...
 1 mov rbx, rdi
 ...
 2 mov qword ptr [rdi], offset off_EDFDA8
 ...
 3 mov rdi, rpb
 4 jnz short loc_6
 5 call InheritedMemberInfoContext::Private::~Private()
 ...
 6 ...
 7 mov rdi, rbx
		...
\end{lstlisting}
\end{footnotesize}
	
	\paragraph{Missing VTables}
	With higher levels of optimization, the compiler removes the entire VTable of any class whose instance is not created irrespective of whether an instance of its base class is created. In Table~\ref{missing_bases}, we counted the total number of classes which constitute the edges not recovered by \codename\ for both libraries and executables. Constructor-destructor as well as a OLA ensure that every information present in the binary is recovered, however, if information is not present in the binary, there is nothing to work with.
	
	\paragraph{Optimized vptr Initialization}
	We found two cases of optimized vptr initialization that the compiler performs. These initializations violate the typical construction/destruction behavior assumed by prior efforts.
	
	\vspace{.06in}
	\noindent
	{\em a. Missing intermediate class initialization}: 
	The below code snippet shows the destructor of class SList\textless MemberList\textgreater in doxygen. 
\begin{footnotesize}
\begin{lstlisting}
 Slist<MemberList>::~Slist()
 ...
 mov rbx, rdi
 #Init vptr of Slist<MemberList>
 mov [rdi], 0xb49d90 
 ...
 mov rdi, rbx
 ...
 #Call dtor of most base class
 call QGList::~QGList() 
 ...
\end{lstlisting}
\end{footnotesize}
	From the ground truth, the direct base class of SList\textless MemberList\textgreater is QList\textless MemberList\textgreater, which inherits from QGList. However, the snippet shows that call to the destructor of QList\textless MemberList\textgreater has been optimized and replaced with that of the most base class. In cases like this, we are unable to identify the direct base class. Table~\ref{false_positives} shows the number of edges with missing intermediate base class. Note that for our evaluation, we neither consider this edges as false positives nor true positives.
	
	\vspace{.06in}
	\noindent
	{\em b. Missing all bases}: The below code snippet shows the only instance of construction of class SPxHarrisRT in Soplex which inherits from SPxRatioTester. 
\begin{footnotesize}
\begin{lstlisting}
 int _cdecl main():
 ...
 call operator new(unsigned long)
 ...
 #Init vptr of soplex::SPxHarrisRT
 mov [rax], 0x457b50 
 ...
\end{lstlisting}
\end{footnotesize}
	However in the binary, only the vptr of SPxHarrisRT gets written into the object address, without initialization of the vptr of SPxRatioTester. Other derived classes of SPxRatioTester were initialized similarly. We found such cases to be common, and leads to inference inaccuracy.

	\section{Related Work}\label{sec:related}
	Multiple prior C++ binary-level solutions have recovered semantic information from a binary~\cite{marx:andre, vci:2017,prakash:2015:vfguard,gawlik:2014:tvip,Dewey:2012:CFI,zhang:2015:vtint,vantough:call:2016}.
	However, VCI~\cite{vci:2017} and Marx~\cite{marx:andre} are the most recent and relevant tools closest to our work. 
	VCI uses constructor only analysis to reconstruct the class inheritance of a program. 
	They handle constructor inlining by relaxing the requirement that the vptr is written into the first argument (implicit this pointer) passed to the function being analyzed. 
	This results in wrong identification of functions as constructors which subsequently result in false inheritance inference. 
	Further, the lack of distinction between non-virtual constructors and destructors also affects inference accuracy.
	
	Andre et al.~\cite{marx:andre} presented Marx which reconstructs class inheritance from binary using certain heuristics. It uses overwrite analysis and groups vptrs written into the same memory location into a single set, only related vptrs get overwritten in the same memory location. 
	Even though Marx is able to correctly group related classes into sets, it does not reason about the direction of inheritance which significantly limits its application.
	
	OBJDigger, presented by Jin et al.~\cite{Jin:2014:objdigger}, uses symbolic execution and inter-procedural data flow analysis to recover object instances, data members and methods of the same class. This is achieved by tracking the usage and propagation of the \textit{this pointer} within and between functions. While the authors did not attempt to recover class inheritance, a method to achieve that was described. However, this can only identify primary base class since they assume that a base class will write its vptr only in the zero offset from the object address. A secondary base class will write to a positive non-zero offset from the object address but that was not accounted for. Also, its virtual table identification approach is weak, it does not take advantage of the well defined structure of a virtual table.
	
	OOAnalyzer~\cite{ooanalyzer:2018} mainly groups methods into classes by combining traditional binary analysis, symbolic analysis and Prolog-based reasoning. The paper explained that class size and VTable size can be considered to decide inheritance. Since OOAnalyzer also considers non-polymorphic classes, one would assume that class size will be relied upon more for this. However, this was not evaluated, therefore, there is no way to confirm the claim that OOAnalyzer can decide inheritance.
	
	Katz et al.\cite{katz:2016} proposed an approach to statically determine the possible targets of virtual function calls. This is achieved by first identifying object tracelets, a statically constructed sequences of operations performed on an object. These object tracelets are then used to train a statistical language model (SLM) for each type. The resulting ensemble of SLMs is used to generate a ranking of their most likely types, from which the likely targets of dynamic dispatches are deduced. Basically, the ensemble of SLMs is used to measure the likelihood that sets of tracelets share the same source, those set of tracelets are grouped together, which then form the basis for predicting possible targets of virtual function calls. The grouping of object types is similar to what Marx does.
	
	Fokin et al.~\cite{fokin2010reconstruction} presented SmartDec which partially recovers certain C++ specific language constructs statically. It attempts to recover classes and their inheritance, virtual and non-virtual member functions, calls to virtual functions, exception raising and handling statements. Its main limitation is the inability to differentiate between between inheritance and composition which results in wrong relationship inference.
	
	Lego~\cite{lego:2013}, presented by Srinivasan et al., uses dynamic analysis to monitor objects allocated at runtime, the lifetime of those objects and methods invoked on them. It uses the set of methods calls involved in the cleanup of an object (basically destructor) to infer inheritance and the depth of these  method calls indicate the number of levels in the class hierarchy of the given object. Lego has two main challenges, 1. the precision of the class inheritance recoverable is limited to the portion of binary that gets invoked during executable, 2. it is unable to differentiate inheritance from composition.
	
	Rewards~\cite{lin:2010:Rewards} is one of the many (e.g., TIE~\cite{lee:2011:Tie}, Laika~\cite{cozzie2008digging}) data structure reverse engineering tools to infer type information from binaries. It uses dynamic analysis to collect and analyze runtime information of a program and then uses that information to recover syntax and semantics of data structures observed during the execution. 
	Rewards only attempts to infer primitive data types of variables and their semantics, it does not attempt to infer high level constructs like objects or the relationships among them.
	
	\section{Conclusion}\label{sec:conclude}
	Extracting class inheritance tree from optimized C++ code is hard, yet useful. We present \codename, a static-analysis based inference engine that employs multiple novel techniques and infers significant amount of directed class inheritance tree from 16 C++ binaries compiled with gcc O0 and O2 options. 
	


\begin{thebibliography}{10}
		\providecommand{\url}[1]{#1}
		\csname url@samestyle\endcsname
		\providecommand{\newblock}{\relax}
		\providecommand{\bibinfo}[2]{#2}
		\providecommand{\BIBentrySTDinterwordspacing}{\spaceskip=0pt\relax}
		\providecommand{\BIBentryALTinterwordstretchfactor}{4}
		\providecommand{\BIBentryALTinterwordspacing}{\spaceskip=\fontdimen2\font plus
			\BIBentryALTinterwordstretchfactor\fontdimen3\font minus
			\fontdimen4\font\relax}
		\providecommand{\BIBforeignlanguage}[2]{{%
				\expandafter\ifx\csname l@#1\endcsname\relax
				\typeout{** WARNING: IEEEtranS.bst: No hyphenation pattern has been}%
				\typeout{** loaded for the language `#1'. Using the pattern for}%
				\typeout{** the default language instead.}%
				\else
				\language=\csname l@#1\endcsname
				\fi
				#2}}
		\providecommand{\BIBdecl}{\relax}
		\BIBdecl
		
		\bibitem{ItaniumABI}
		``{Itanium C++ ABI},'' http://refspecs.linuxbase.org/cxxabi-1.83.html,
		Revision: 1.83.
		
		\bibitem{lego:2013}
		T.~{ Reps} and V.~K. Srinivasan, ``{Software-Architecture Recovery from Machine
			Code},'' http://digital.library.wisc.edu/1793/65091, Tech. Rep. TR1781, March
		2013.
		
		\bibitem{bounov:2016:ivt}
		D.~Bounov, R.~G. K{\i}c{\i}, and S.~Lerner, ``{Protecting C++ dynamic dispatch
			through vtable interleaving},'' in \emph{{Proceedings of the 23rd Annual
				Network and Distributed System Security Symposium}}, ser. {NDSS {\rq}16}, Feb
		2016.
		
		\bibitem{Brumley:BAP}
		\BIBentryALTinterwordspacing
		D.~Brumley, I.~Jager, T.~Avgerinos, and E.~J. Schwartz, ``{BAP: A Binary
			Analysis Platform},'' in \emph{{Proceedings of the 23rd International
				Conference on Computer Aided Verification}}, ser. {CAV'11}.\hskip 1em plus
		0.5em minus 0.4em\relax Berlin, Heidelberg: Springer-Verlag, 2011, pp.
		463--469. [Online]. Available:
		\url{http://dl.acm.org/citation.cfm?id=2032305.2032342}
		\BIBentrySTDinterwordspacing
		
		\bibitem{cozzie2008digging}
		A.~Cozzie, F.~Stratton, H.~Xue, and S.~T. King, ``{Digging for Data
			Structures.}'' in \emph{{OSDI}}, vol.~8, 2008, pp. 255--266.
		
		\bibitem{Dewey:2012:CFI}
		D.~Dewey and J.~T. Giffin, ``{Static detection of {C++} vtable escape
			vulnerabilities in binary code.}'' in \emph{{Proceedings of 19th Annual
				Network and Distributed System Security Symposium (NDSS'12)}}, 2012.
		
		\bibitem{vci:2017}
		\BIBentryALTinterwordspacing
		M.~Elsabagh, D.~Fleck, and A.~Stavrou, ``{Strict Virtual Call Integrity
			Checking for C++ Binaries},'' in \emph{{Proceedings of the 2017 ACM on Asia
				Conference on Computer and Communications Security}}, ser. {ASIA CCS
			'17}.\hskip 1em plus 0.5em minus 0.4em\relax New York, NY, USA: ACM, 2017,
		pp. 140--154. [Online]. Available:
		\url{http://doi.acm.org/10.1145/3052973.3052976}
		\BIBentrySTDinterwordspacing
		
		\bibitem{fokin:2011:C++Rev}
		A.~Fokin, E.~Derevenetc, A.~Chernov, and K.~Troshina, ``{SmartDec: Approaching
			{C}++ Decompilation},'' in \emph{{Reverse Engineering (WCRE), 2011 18th
				Working Conference on}}, 2011, pp. 347--356.
		
		\bibitem{fokin2010reconstruction}
		A.~Fokin, K.~Troshina, and A.~Chernov, ``{Reconstruction of class hierarchies
			for decompilation of C++ programs},'' in \emph{{Software Maintenance and
				Reengineering (CSMR), 2010 14th European Conference on}}.\hskip 1em plus
		0.5em minus 0.4em\relax IEEE, 2010, pp. 240--243.
		
		\bibitem{gawlik:2014:tvip}
		R.~Gawlik and T.~Holz, ``{Towards Automated Integrity Protection of {C}++
			Virtual Function Tables in Binary Programs},'' in \emph{{Proceedings of 30th
				Annual Computer Security Applications Conference (ACSAC'14)}}, Dec 2014.
		
		\bibitem{haller2015shrinkwrap}
		I.~Haller, E.~G{\"o}kta\c{s}, E.~Athanasopoulos, G.~Portokalidis, and H.~Bos,
		``{ShrinkWrap: VTable Protection without Loose Ends},'' in \emph{{Proceedings
				of the 31st Annual Computer Security Applications Conference}}.\hskip 1em
		plus 0.5em minus 0.4em\relax ACM, 2015, pp. 341--350.
		
		\bibitem{jang:2014:CFI:Src}
		D.~Jang, Z.~Tatlock, and S.~Lerner, ``{Safe{D}ispatch: Securing {C++} Virtual
			Calls from Memory Corruption Attacks},'' in \emph{{Proceedings of 21st Annual
				Network and Distributed System Security Symposium (NDSS'14)}}, 2014.
		
		\bibitem{Jin:2014:objdigger}
		W.~Jin, C.~Cohen, J.~Gennari, C.~Hines, S.~Chaki, A.~Gurfinkel, J.~Havrilla,
		and P.~Narasimhan, ``{Recovering {C}++ Objects From Binaries Using
			Inter-Procedural Data-Flow Analysis},'' in \emph{{Proceedings of ACM SIGPLAN
				on Program Protection and Reverse Engineering Workshop (PPREW'14)}}, 2014,
		pp. 1--11.
		
		\bibitem{katz:2016}
		O.~Katz, R.~El-Yaniv, and E.~Yahav, ``{Estimating Types in Binaries using
			Predictive Modeling},'' in \emph{{Proceedings of the 43rd Annual ACM
				SIGPLAN-SIGACT Symposium on Principles of Programming Languages}}, ser. {POPL
			'16}.\hskip 1em plus 0.5em minus 0.4em\relax New York, NY, USA: ACM, Jan
		2016, pp. 313--326.
		
		\bibitem{lee:2015:caver}
		B.~Lee, C.~Song, T.~Kim, and W.~Lee, ``{Type casting verification: Stopping an
			emerging attack vector},'' in \emph{{24th USENIX Security Symposium (USENIX
				Security 15)}}, 2015, pp. 81--96.
		
		\bibitem{lee:2011:Tie}
		J.~H. Lee, T.~Avgerinos, and D.~Brumley, ``{TIE: Principled Reverse Engineering
			of Types in Binary Programs},'' in \emph{{Proceedings of the 18th Annual
				Network and Distributed System Security Symposium (NDSS'11)}}, 2011.
		
		\bibitem{lin:2010:Rewards}
		Z.~Lin, X.~Zhang, and D.~Xu, ``{Automatic Reverse Engineering of Data
			Structures from Binary Execution},'' in \emph{{Proceedings of the 17th Annual
				Network and Distributed System Security Symposium (NDSS'10)}}, 2010.
		
		\bibitem{Burow2017CFIXXOT}
		{Nathan Burow and Derrick McKee and Scott A. Carr and Mathias Payer}, ``{CFIXX:
			Object Type Integrity for C++ Virtual Dispatch},'' in \emph{{Proceedings of
				the 25th Annual Network and Distributed System Security Symposium}}, ser.
		{NDSS'18}, San Diego, CA, USA, {2018}.
		
		\bibitem{marx:andre}
		A.~Pawlowski, M.~Contag, V.~van~der Veen, C.~Ouwehand, T.~Holz, H.~Bos,
		E.~Athanasopoulos, and C.~Giuffrida, ``{MARX : Uncovering Class Hierarchies
			in C++ Programs},'' in \emph{{Proceedings of the 24th Annual Network and
				Distributed System Security Symposium}}, ser. {NDSS '17}, San Diego, CA, USA,
		2017, p.~15.
		
		\bibitem{prakash:2015:vfguard}
		A.~Prakash, X.~Hu, and H.~Yin, ``{vfGuard: Strict Protection for Virtual
			Function Calls in COTS C++ Binaries},'' in \emph{{Proceedings of the 22nd
				Annual Network and Distributed System Security Symposium (NDSS'15)}}, 2015.
		
		\bibitem{Ray:1994:MSVC}
		J.~Ray, ``{C++: Under the Hood},''
		http://www.openrce.org/articles/files/jangrayhood.pdf, March 1994.
		
		\bibitem{Sabanal:2007:CppRev}
		P.~V. Sabanal and M.~V. Yason, ``{Reversing {C}++},'' \emph{Blackhat Security
			Conference}, 2007.
		
		\bibitem{ooanalyzer:2018}
		E.~J. Schwartz, C.~F. Cohen, M.~Duggan, J.~Gennari, J.~S. Havrilla, and
		C.~Hines, ``{ Using Logic Programming to Recover C++ Classes and Methods from
			Compiled Executables},'' in \emph{{2018 ACM SIGSAC Conference on Computer and
				Communications Security}}, ser. {CCS {\rq}18}.\hskip 1em plus 0.5em minus
		0.4em\relax New York, NY, USA,: ACM, October 15--19 2018, p.~16.
		
		\bibitem{igor:ida:decompiler}
		\BIBentryALTinterwordspacing
		I.~Skochinsky, ``{Practical {C}++ decompilation},'' 2011. [Online]. Available:
		\url{https://archive.org/details/Recon\_2011\_Practical\_Cpp\_decompilation}
		\BIBentrySTDinterwordspacing
		
		\bibitem{srinivasan2013software}
		V.~K. Srinivasan and T.~Reps, ``{Software Architecture Recovery from Machine
			Code},'' Technical Report TR1781, University of Wisconsin-Madison, Tech.
		Rep., 2013.
		
		\bibitem{vtvgcc:2012:tice}
		C.~Tice, T.~Roeder, P.~Collingbourne, S.~Checkoway, {\'U}.~Erlingsson,
		L.~Lozano, and G.~Pike, ``{Enforcing Forward-Edge Control-Flow Integrity in
			{GCC} \& {LLVM}},'' in \emph{{Proceedings of 23rd USENIX Security Symposium
				(USENIX Security'14)}}, 2014, pp. 941--955.
		
		\bibitem{vantough:call:2016}
		V.~van~der Veen, E.~G{\"o}ktas, M.~Contag, A.~Pawlowski, X.~Chen, S.~Rawat,
		H.~Bos, T.~Holz, E.~Athanasopoulos, and C.~Giuffrida, ``{A Tough call:
			Mitigating Advanced Code-Reuse Attacks At The Binary Level},'' in
		\emph{{Proceedings of IEEE Symposium on Security and Privacy}}, ser. {Oakland
			{\rq}16}, May 2016.
		
		\bibitem{zhang2016vtrust}
		C.~Zhang, S.~A. Carr, T.~Li, Y.~Ding, C.~Song, M.~Payer, and D.~Song,
		``{VTrust: Regaining Trust on Virtual Calls},'' in \emph{{Network and
				Distributed System Security Symposium}}, 2016.
		
		\bibitem{zhang:2015:vtint}
		C.~Zhang, C.~Song, Z.~K. Chen, Z.~Chen, and D.~Song, ``{{VTint:} Defending
			Virtual Function Tables' Integrity},'' in \emph{{Proceedings of the 22nd
				Annual Network and Distributed System Security Symposium (NDSS'15)}}, 2015.
		
	\end{thebibliography}
\end{document}